\documentclass[aps,pra,superscriptaddress,twocolumn]{revtex4}
\usepackage[utf8x]{inputenc}
\usepackage{float}
\usepackage{bm}
\usepackage{graphicx,amsmath,latexsym,amssymb}
\usepackage{color}
\usepackage{dcolumn}

\def\bra#1{\mathinner{\langle{#1}|}}
\def\ket#1{\mathinner{|{#1}\rangle}}

\def\Bra#1{\left<1>}
{\catcode`\|=\active\gdef\Braket#1{\left<\mathcode`\|"8000\let|\bravert {#1}\right>}}
\def\bravert{\egroup\,\vrule\,\bgroup}

\begin{document}
\title{Optical response of a binary atomic system with incoherent gain}
\author{L. Acevedo}
%\email{manuel.donaire@uva.es}
\affiliation{Departamento de F\'isica Te\'orica, At\'omica y \'Optica,  Universidad de Valladolid, Paseo Bel\'en 7, 47011 Valladolid, Spain}
\author{J. S\'anchez-C\'anovas}
%\email{manuel.donaire@uva.es}
\affiliation{Departamento de F\'isica Te\'orica, At\'omica y \'Optica,  Universidad de Valladolid, Paseo Bel\'en 7, 47011 Valladolid, Spain}
\author{M. Donaire}
\email{manuel.donaire@uva.es}
\affiliation{Departamento de F\'isica Te\'orica, At\'omica y \'Optica,  Universidad de Valladolid, Paseo Bel\'en 7, 47011 Valladolid, Spain}

%\author{J. S\'anchez-C\'anovas}
%\email{manuel.donaire@uva.es}
%\affiliation{Departamento de F\'isica Te\'orica, At\'omica y \'Optica,  Universidad de Valladolid, Paseo Bel\'en 7, 47011 Valladolid, Spain}
%\author{M. Donaire}
%\email{manuel.donaire@uva.es}
%\affiliation{Departamento de F\'isica Te\'orica, At\'omica y \'Optica,  Universidad de Valladolid, Paseo Bel\'en 7, 47011 Valladolid, Spain}

\begin{abstract}
We study the optical response of a binary system of identical atoms in which one of them is excited by an incoherent pump.  %We study the properties of photon scattering, absorption and emission, together with the time evolution of the atomic system. %This allows us to characterize and eventually manipulate the state of the system, paving the way for prospective applications in quantum information processing. 
Applying the diagrammatic formalism developed in  Donaire [Phys. Rev. A\textbf{104} 043704 (2021)], it is shown how scattering, absorption, stimulated emission, spontaneous emission and resonant energy transfer can be tailored by varying i) the interatomic distance, which governs the interference effects of the emitted radiation; and ii) the pump rate, which determines the population of the atomic levels. It is found that,  for sufficiently strong pumping, the collective component of the extinction cross-section becomes negligible,  regardless of the interatomic distance, as optical gains compensate for losses, and the total extinction cross-section is reduced to less than  half of its value in the absence of pumping. In contrast, at weak pumping and short interatomic distances, interference effects lead to a significant suppression of the extinction cross-section relative to that of two noninteracting atoms. %For weak pump, in contrast, it is as a result of interference and for short interatomic distances that the total extinction cross-section becomes considerably smaller than its value for two isolated atoms. % collective extinction cross-section presents a remarkable asymmetry with respect to the detuning, becoming highly negative at resonance.  %However, in contrast to the result of semiclassical approaches, the total extinction cross-section does not vanish, regardless of the interatomic distance and the pump rate. 
\end{abstract}

\maketitle

\section{Introduction and Motivation}

The optical response of a medium made of point dipoles has been extensively studied in different scenarios of physical interest. These include random media of the sort of molecular and atomic gases \cite{Shalaev,vanTiggelen,vanTiggelen2}, highly correlated solids of the kind of liquid crystals \cite{Petrov}, and artificial solids like photonic crystals \cite{Sakoda}. In all of them, the features of interest are related to the collective and nonlocal optical response of the media in their ground state. Besides the study of global properties like scattering patterns and transmittance of light, the study of the local density of states through the fluorescence  of a single excited molecule embedded within a medium of molecules are performed with all the molecules in their internal ground state \cite{myPRA1}.

For the case of excited systems, the characterization of their excited states is accomplished through the study of their emission properties. Thus, in an ensemble of identical atoms which share excited states, like in Dicke's problem \cite{Dicke}, the interest has concentrated on the superradiance and subradiance properties. In the simpler case of two two-level atoms sharing a single excitation, superradiant-symmetric and  subradiant-antisymmetric states arise when the atoms are sufficiently close to each other \cite{Craigbook}. Closely related, in the context of quantum computing, coherence times and emission properties are also features of interest for the encoding and read-out of information using atomic qubits \cite{Broadway}.  Hence, periodic arrays of atoms present a fertile ground for the exploration of collective and coherent optical phenomena in structured media. In them, the interplay of lattice periodicity and dipolar interactions leads to the emergence of lattice resonances, which can drastically enhance light-matter coupling and give rise to collective modes with narrow linewidths and long lifetimes \cite{la19}. When the arrays are coherently driven, strong light-matter coupling manifests in features such as the Mollow triplet, which has been observed in atomic lattices in regimes of strong driving and coherent scattering \cite{la21}. These structured systems also enable photon-mediated interactions across the array, leading to phenomena like photon control, non-reciprocal transport, and collective excitation dynamics \cite{la22}. Also for practical purposes, in the fields of sensors \cite{sensors}, nanoantennas \cite{antenas}, photovoltaics \cite{photovolts}, light-emitting devices, etc., the manipulation of the scattering pattern of ensembles of dipoles with active elements has inspired recent research activities \cite{Vincenzo,scatotros}. In these non-Hermitian systems gains and losses are implemented in a way to inhibit scattering, %at least in some directions, 
and certain configurations may lead to responses compatible with PT-symmetric systems. This is for instance the case of classical dimers with balanced gains and losses \cite{ManjavacasPT1,ChenetalPRA,ManjavacasSandersPT2}, where a classical approach has been applied to the optical response of metallic and dielectric nanoparticles with gains.

Motivated by the aforementioned interest in monitoring the state evolution and the optical properties of excited systems of point dipoles, it is our purpose to perform a fully quantum study of the optical response of  excited atomic systems. Our study aims at unrevealing the impact of the manner 
 in which the systems are excited as well as of the spatial correlations of their constituents in their collective and nonlocal optical responses. Concerning the manner in which the systems are excited,  we will distinguish between coherent and incoherent excitation. Also, we will differentiate between stationary and transient optical responses. In regards to the spatial correlations, we will distinguish between regular and random configurations.  The resultant study is meant to pave the way for the experimental detection of interatomic forces through scattering, and for the manipulation and read-out of quantum information processing in atomic qubits.

 Within this framework, the present article is the second one of a series devoted to the characterization of the optical response of excited atomic systems. In particular,  following up with the starting article of Ref.\cite{myPRA}, in which one of us has studied the optical response of a single atom incoherently excited, and inspired by the aforementioned  research  on dimers with active elements, in this article we characterize the optical response of a binary system of identical atoms, with one of them excited by an incoherent pump, and continuously illuminated by a weak and linearly polarized probe field. 
%Thus, it is one the aims of this article to perform an analogous study on a binary atomic system, employing a fully quantum approach. Special attention is to be paid to the chiral properties of the response, as well as to the time evolution of the atomic system. In addition, 
%Our study is meant to pave the way for further investigations in two-atom qubits for prospective applications in quantum information processing.   In particular,  the increase of coherence times and the efficient read out of the atomic states are desirable features in entangled systems. Motivated by all this, in this article we characterize the optical response of a binary system of identical atoms, with one of them excited by an incoherent pump, and continuously illuminated by a weak and polarized probe field. General statements are derived and comparison is made with previous classical results in analogous systems. In particular we will study the anomalies already 
%found in the optical response of analogous systems in relation to $\mathcal{PT}$-symmetry \cite{ManjavacasPT,Khandekar,JKP16}.
From a fundamental perspective, since the use of classical polarizability models was found to be insufficient to describe the optical response of excited atoms, we exploit the quantum diagrammatic-wavefunction approach developed in Ref.\cite{myPRA} to track the dynamics of both atomic states and photons, to identify and  classify all the radiative processes, and to compute the cross-section and emission power.  From a practical perspective, we will show that losses are partially compensated, either by stimulated emission or as a result of scattering interference for certain interatomic distance. Hence, although the collective extinction cross-section may vanish as a result of the combination of gains and interference effects,  the total extinction cross-section stays positive regardless of the pump rate and the interatomic distance. This is in contrast to the results of classical dimers \cite{ManjavacasPT1,ManjavacasSandersPT2,Japs,ORN19,JKP16}.

% on the optical response 
%of a three-level atom to a quasi-resonant probe field while subjected to incoherent pumping. In that approach,  the incoherent dynamics caused by the pump is treated in an effective manner that is compatible with unitarity. It allows one to track the dynamics of both atomic states and photons, to identify and  classify all the radiative processes, and to compute the cross-section and emission power. In Ref.\cite{myPRA} it was shown that, for sufficiently strong pump, gains and losses compensate due to stimulated emission, resulting in zero total extinction cross-section. From a more fundamental perspective, it was shown that the use of classical polarizability models are insufficient to describe the optical response of excited atoms, and a fully quantum approach becomes necessary. Here, for the case of a binary system,  we will 
%see that a semiclassical approach is even insufficient to describe the optical response of a binary system of atoms in their ground state. 

%In this respect, in a subsequent publication we will study the asymmetric response of the system with respect to the side of incidence of the probe field, and a comparative study will be performed with respect to the asymmetric response of classical dimers.  

In prospective works,  we will study the asymmetric response of the same system with respect to the side of incidence of the probe field, and we will analyze the optical response of single atoms and binary atomic systems coherently excited. Further, we will address these problems in ensembles of many atoms, both in regular and random configurations. %For practical properties, special attention will be paid to the chiral properties of the response, as well as to the possibility to monitor the states of the systems by photon scattering. %Our study is meant to pave the way for the experimental detection of interatomic forces through scattering, and for the manipulation and read-out of quantum information processing in atomic qubits.

The article is organized as follows. In Sec.~\ref{lasec2} we outline the fundamentals of the formalism. Sec.~\ref{lasec3} contains the classification of all the radiative processes together with the calculation of emission powers and cross-sections. We finalize with the discussion of the results in Sec.~\ref{lasec4}.

%\textcolor{red}{IMPORTANTE: HACER LA COMPARACIÓN DEL RESULTADO DE LAS SECCIONES EFICACES A CERO PUMP CON EL SCATTERING DE DIMEROS ATOMICOS EN FORMALISMO SEMICLASICO, E.G., EL DE VAN TIGEELEN. EXISTE DE POR SI UNA DIFERENCIA CUALITATIVA CONCERNIENTE A LA ASIMETRÍA CON RESPECTO AL DETUNNING. Y CREO QUE EL RESTO DE DIFERENCIAS TIENEN QUE VER CON QUE NUESTRO CÁLCULO INLCUYE ESTADOS A DOS FOTONES (I.E., DEL PROBE FIELD Y VIRTUALES), UNA ESPECIA DE NO-LINEALIDAD, QUE NO ESTÁ PRESENTE EN EL FORMALISMO LINEAL SEMICLASICO DONDE INDUCCIONES Y EMISIONES SON CONSECUTIVAS CON $G^{0}$S.}

\section{Fundamentals}\label{lasec2}
\begin{figure}
\begin{center}
\includegraphics[height=30mm,width=80mm,angle=0,clip]{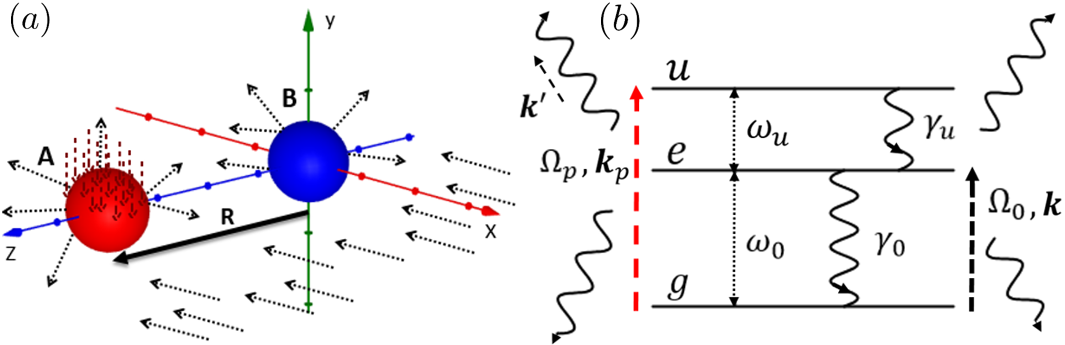}
\caption{Schematics of the system under study, consisting of two identical three-level atoms and two external fields. A pump field of strength $\Omega_{p}$ causes the transient excitation of atom $A$ from its ground state $g$ to its upper level $u$, from which it decays, rapidly and incoherently, to the intermediate level $e$ at a rate 
$\gamma_{u}$, causing an effective pump rate $\mathcal{P}=\Omega_{p}^{2}/\gamma_{u}$. Energy intervals and dissipative channels are depicted. Atom $B$ is in its ground state. The two atoms are illuminated by a weak probe field of strength 
$\Omega_{0}$ and momentum \textbf{k}, quasi-resonant 
with the $g\rightarrow e$ transition. Radiation of momentum $\bf{k'}$ is scattered.} \label{fig1}
\end{center}
\end{figure}

The two atoms of our system are modelled as three-level ones. The ground state $g$ is  unique, the first excited state $e$ is metastable and degenerate, with lifetime $\gamma_{0}^{-1}$, and the upper auxiliary state $u$ is highly unstable, decaying into $e$ in a lifetime $\gamma_{u}^{-1}$ which satisfies $\gamma_{u}^{-1}\ll\gamma_{0}^{-1}$. Their respective frequency transitions are $\omega_{0}$ and $\omega_{u}$, respectively.  Whereas only the active atom is incoherently pumped by a strong pump field quasi-resonant with the transition $g\rightleftarrows u$, the two atoms are illuminated with a weak probe field which is quasi-resonant with one of the transition  $g\rightleftarrows e$ --see Fig.\ref{fig1}. The amplitudes, frequencies, momenta and polarization vectors of these fields are $E_{0}$, $\omega$, $\mathbf{k}$, $\bm{\epsilon}$, 
and $E_{p}$, $\omega_{p}$, $\mathbf{k}_{p}$, $\bm{\epsilon}_{p}$, respectively. Their corresponding 
 Rabi frequencies are $\Omega_{0}=E_{0}\boldsymbol{\mu}\cdot\bm{\epsilon}/\hbar$ and 
$\Omega_{p}=E_{p}\tilde{\boldsymbol{\mu}}\cdot\bm{\epsilon}_{p}/\hbar$, with $\boldsymbol{\mu}=\langle g|\mathbf{d}|e\rangle$ and 
$\tilde{\boldsymbol{\mu}}=\langle g|\mathbf{d}|u\rangle$ being the dipole transition moments, and $\mathbf{d}$ being the electric dipole operator. For further purposes we will assume that  $\boldsymbol{\mu}$ is  axially symmetric with respect to the interatomic axis. In turn, this implies an equal contribution of near field and far field interatomic interactions to the optical response of the system. The probe field interacts weakly with the atoms, i.e., $\Omega_{0}\ll\gamma_{0}$, whereas incoherent pumping is achieved for $\gamma_{u}\gg\Omega_{p},\gamma_{0}$, in which case the fast dynamics of the auxiliary state can be integrated out in an effective 
manner. The incoherent pump implemented this way guarantees the existence of a steady state made of a statistical (i.e., incoherent) mixture of the populations $g$ and $e$ in the active atom. 

Following Refs.\cite{German,myPRA}, after integrating out the dynamics of the auxiliary state $u$, the resultant Bloch equations for the density matrix of the effective two-level atom read % we read that the pump attenuates the coherent evolution between the two levels $g$ and $e$ in the pumped atom. Defining $\Gamma=\gamma+\mathcal{P}$, straightforward integration of the above equations leads to the solutions
%$\Omega=\mathbf{d}\cdot\mathbf{E}_{0}/2$, 
%\subsection{Derivation of the single-atom polarizability and optical cross-section}\label{lasec2b}
\begin{align}\label{populations}
\rho_{ee}(t)&=\frac{\mathcal{P}}{\Gamma}(1-e^{-\Gamma t})+N_{e}e^{-\Gamma t},\\
\rho_{gg}(t)&=\frac{\gamma}{\Gamma}(1-e^{-\Gamma t})+e^{-\Gamma t}(1-N_{e}),
\end{align}
where $\gamma$ is the total decay rate of the transition $e\rightarrow g$, sum of the natural width $\gamma_{0}$ and the non-radiative decay rate $\gamma_{nr}$; $\mathcal{P}$ is the pump rate, $\mathcal{P}=\Omega_{p}^{2}/\gamma_{u}$, and $\Gamma=\gamma+\mathcal{P}$. $\rho_{gg}$ and $\rho_{ee}$ are the level populations of the active atom,  with $N_{e}$ being the excited population of the active atom at $t=0$. From the coherence element $\rho_{eg}$ we read that the pump attenuates the coherent evolution between the two levels. In our wavefunction formalism  this element  will provide, in an effective manner, the time attenuating factors  associated to incoherence, whether to gains or to losses.  For asymptotic times, $\Gamma t\gg1$, the atomic populations converge to the 
steady values $\rho_{gg}(t\rightarrow\infty)=\gamma/\Gamma$, $\rho_{ee}(t\rightarrow\infty)=\mathcal{P}/\Gamma$, while the coherence element vanishes at a rate $\Gamma$ regardless of the initial conditions.

Thus, we can write the steady state atoms-EM field as a mixed state made of the incoherent superposition of the pure states
\begin{equation}
	%|\Psi_{0}(\tau)\rangle&=|\Psi_{0}(\tau)\rangle_{g}\oplus|\Psi_{0}(\tau)\rangle_{e},\nonumber\\
	|\Psi_{0}\rangle_{g}=\sqrt{\gamma/\Gamma}|N_{\mathbf{k},\bm{\epsilon}};\tilde{g},g\rangle,\quad
	|\Psi_{0}\rangle_{e}=\sqrt{\mathcal{P}/\Gamma}|N_{\mathbf{k},\bm{\epsilon}};\tilde{e},g\rangle.
	\label{wavefunction}
\end{equation}
In these equations the 'tilde' states refer to the active atom, and $|N_{\mathbf{k},\bm{\epsilon}}\rangle$ is the EM state which contains the $N_{\mathbf{k},\bm{\epsilon}}$ photons of the probe field,
\begin{equation}
	|N_{\mathbf{k},\bm{\epsilon}}\rangle=\frac{1}{\sqrt{N_{\mathbf{k},\bm{\epsilon}}!}}
	\left(a_{\mathbf{k},\bm{\epsilon}}^{\dagger}\right)^{N_{\mathbf{k},\bm{\epsilon}}}\ket{0},\nonumber
\end{equation}
where $N_{\mathbf{k},\bm{\epsilon}}/\mathcal{V}=\epsilon_{0}E_{0}^{2}/2\hbar\omega$ is the photon density and  $\epsilon_{0}c\!E_{0}^{2}/2$ is the time-averaged intensity.
The atomic states, with labels $A$ and $B$, are accompanied by the statistical weights of the populations of the active atom, $\sqrt{\gamma/\Gamma}$ and $\sqrt{\mathcal{P}/\Gamma}$, respectively. 
It is upon the statistical mixture of these states that quantum perturbation theory is to be applied in the computation of the optical response. Thus, the physical 
quantities to be calculated are statistical averages over the quantum expectation values computed upon the pure states $|\Psi_{0}\rangle_{g}$ and 
$|\Psi_{0}\rangle_{e}$.

In regards to the atomic interactions with the probe field,  we employ time-dependent perturbation theory using the wavefunction formalism. It is based on the time propagator of the atoms-EM field system, $\mathbb{U}(t)$. In terms of the 
Hamiltonian of the system,  $H$ , it reads \cite{Sakurai}
\begin{equation}
\mathbb{U}(t-t_{0})=\mathcal{T}\textrm{-exp}\left\{-i\hbar^{-1}\int_{t_{0}}^{t}d\tau\:H(\tau)\right\},
\end{equation}
where $H$ contains a free component, $H_{0}$, and an interaction term, $W$. As for the free Hamiltonian, after integrating out the auxiliary state $u$, it reads
\begin{equation}
H_{0}=\sum_{i=A,B}\hbar\omega_{0}\ket{e}_{i}\bra{e}_{i}+
\sum_{\mathbf{k}',\bm{\epsilon}'}\hbar\omega'(a^{\dagger}_{\mathbf{k}',\bm{\epsilon}'}a_{\mathbf{k}',\bm{\epsilon}'}+\frac{1}{2}),\nonumber 
\end{equation}
where the script $i$ denotes the atom, and the second term corresponds to the free EM Hamiltonian, with $a_{\mathbf{k}',\bm{\epsilon}'}^{\dagger}$ and 
$a_{\mathbf{k}',\bm{\epsilon}'}$ being the creation and annihilation operators of photons of momentum  $\mathbf{k}'$, frequency $\omega'=c\:k'$ and polarization vector $\bm{\epsilon}'$. 
The atom-field interaction is, in the electric dipole approximation,
\begin{equation}
W=-\sum_{i=A,B}\mathbf{d}_{i}\cdot\mathbf{E}(\mathbf{r}_{i}),\nonumber
\end{equation}
where $\mathbf{r}_{i}$ is the center of mass of each atom. In Schr\"odinger's picture, 
the electric field operator can be expanded as a sum over normal modes \cite{Sakurai,Milonni_book},
\begin{align}
\mathbf{E}(\mathbf{r})&=i\sum_{\mathbf{k}',\bm{\epsilon}'}\sqrt{\frac{\hbar\omega'}{2\epsilon_{0}\mathcal{V}}}
\left[\bm{\epsilon}'a_{\mathbf{k}',\bm{\epsilon}'}e^{i\mathbf{k}'\cdot\mathbf{r}}-\bm{\epsilon}'^{\ast}a^{\dagger}_{\mathbf{k}',\bm{\epsilon}'}
e^{-i\mathbf{k}'\cdot\mathbf{r}}\right]\nonumber\\
&=\sum_{\mathbf{k}',\bm{\epsilon}'}\left[\mathbf{E}^{(+)}_{\mathbf{k}',\bm{\epsilon}'}(\mathbf{r})+\mathbf{E}^{(-)}_{\mathbf{k}',\bm{\epsilon}'}
(\mathbf{r})\right].\label{fieldE}
\end{align}
Essential in our calculations is the vacuum expectation value of the quadratic fluctuations of the electric field, which reads
\begin{equation}
\sum_{\bm{\epsilon}'}\int_{0}^{4\pi}\frac{d\Theta_{\mathbf{k}'}}{8\pi^{2}}\bra{0}\mathbf{E}^{(+)}_{\mathbf{k}',\bm{\epsilon}'}(\mathbf{r})
\mathbf{E}^{(-)}_{\mathbf{k}',\bm{\epsilon}'}(\mathbf{r}')\ket{0}=\frac{-\hbar\omega}{\epsilon_{0}c^{2}}
\textrm{Im}\mathbb{G}(\mathbf{r}-\mathbf{r}';\omega').\nonumber
\end{equation}
Here,  $\mathbb{G} (\mathbf{r}-\mathbf{r}';\omega')$ is the dyadic Green's function of the electric field induced at $\mathbf{r}$ by an electric dipole of 
frequency $\omega'$ located at $\mathbf{r}'$, 
\begin{equation}
\mathbb{G} (\mathbf{R};\omega')=-\frac{k' e^{ik'R}}{4\pi}\left[\frac{\mathbb{P}}{k'R}+\frac{i\mathbb{Q}}{(k'R)^{2}}
-\frac{\mathbb{Q}}{(k'R)^{3}}\right],\label{Green}
\end{equation}
where the tensors $\mathbb{P}$ and $\mathbb{Q}$ read $\mathbb{P}=\mathbb{I}-\mathbf{R}\mathbf{R}/R^{2}$,  $\mathbb{Q}=\mathbb{I}-3\mathbf{R}\mathbf{R}/R^{2}$, 
with $\mathbf{R}=\mathbf{r}-\mathbf{r}'$, $k'=\omega'/c$.

Considering $W$ as a perturbation to $H_{0}$, the time propagator of the system admits an expansion in powers of  $W$ which can be developed from its 
time-ordered exponential expression,
\begin{equation}
\mathbb{U}(t-t_{0})=\mathbb{U}_{0}(t)\:\mathcal{T}\textrm{-exp}\int_{t_{0}}^{t} -i\hbar^{-1}\mathop{d\tau} \mathbb{U}_{0}^{\dagger}(\tau)W
\mathbb{U}_{0}(\tau-t_{0}),\label{propagator}
\end{equation}
where $\mathbb{U}_{0}(t-t')$ is the unperturbed time-propagator, $\mathbb{U}_{0}(t-t')=\exp{[-i\:\hbar^{-1}H_{0}(t-t')]}$. 

For a weak probe field, $\Omega_{0}\ll\Gamma$, the optical response of the atom at leading order involves terms of up to $\mathcal{O}(W^{4})$ in 
$\mathbb{U}$ in Eq.(\ref{propagator}). In all the processes --see, eg., the diagrams of Fig.\ref{fig2},  two of the interaction vertices, $W$, create or annihilate one probe field photon each, whereas the other pair creates and annihilate a virtual photon. For the case of the collective response, one of the vertices of these pairs applies to each atom.  

Finally, an essential point in our wavefunction formalism is the effective implementation of incoherent effects in the intermediate processes within  $\mathbb{U}$ in a way that it results compatible with unitarity. Following Ref.\cite{myPRA}, from Bloch's equations for the atomic dynamics, we read that the coherent transitions from steady to intermediate states of the active atom get attenuated in time at an effective rate $\Gamma/2$. Likewise, for the case of the passive atom, the attenuating factor is $\gamma/2$. Mathematically this implies that exponentially attenuating factors of the sort of $e^{-\Gamma\tau/2}$ and $e^{-\gamma\tau/2}$ are to be implemented in each case for the time evolution of those intermediate states different to the original ones in each process.

\section{Identification of collective radiative processes: Power and cross-sections}\label{lasec3}

In Ref.\cite{myPRA} all the radiative processes involving a single active atom have been identified. The distinctive features of scattering, absorption, stimulated emission and spontaneous emission  processes have been spotted based on a diagrammatic representation. This is in contrast to standard approaches based on linear response theory and density matrix formalism \cite{Carmichel,Scully,Harry_Paul,Hertel-Schulz}. Here we extend the analysis to our binary system, in which collective radiative processes arise when the two atoms get correlated through the exchange of a virtual photon. We consider the system  in the steady state of Eq.(\ref{wavefunction}), continuously illuminated by the 
 weak and quasi-resonant probe field of strength $\Omega_{0}$ and frequency $\omega$ as outlined in Sec.~\ref{lasec2}. We will identify  all the radiative and non-radiative collective processes and represent 
 diagrammatically their probabilities, $P_{n}(t)=|\langle\Psi_{n}^{f}|\Psi_{n}(t)\rangle|^{2}$.  
 In this expression the state $|\Psi_{n}(t)\rangle$ results from the evolution of one of the pure states, $|\Psi_{0}\rangle_{g}$ or $|\Psi_{0}\rangle_{e}$, 
 in a time interval $t$, $|\Psi_{n}(t)\rangle=\mathbb{U}(t)|\Psi_{0}\rangle_{g/e}$; and the state $|\Psi_{n}^{f}\rangle$ is that whose radiative content is 
 to be computed for the calculation of the radiative power, cross-section, etc. The corresponding processes are labeled with the script $n$. Since the illumination is continuous, we will assume the asymptotic condition $\Gamma t\gg1$.\\
 
\subsection{Collective scattering}\label{scatt}
\begin{figure}
	\begin{center}
		\includegraphics[width=82mm,angle=0,clip]{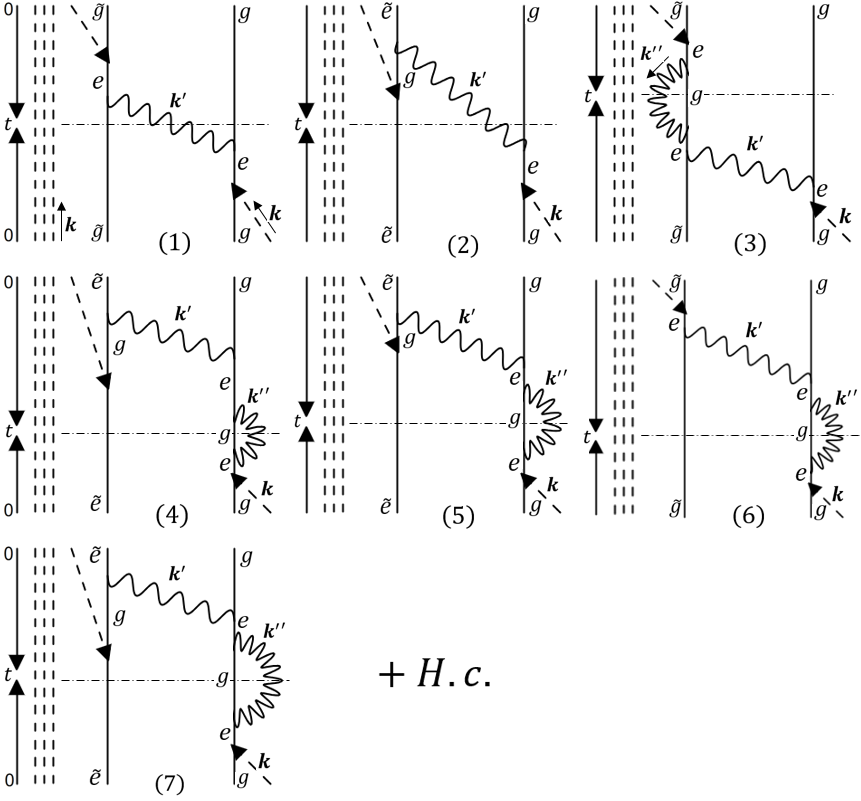}
		\caption{Diagrammatic representation of the collective scattering processes.  Solid straight lines represent atomic states. Wavy lines stand for virtual photons created and annihilated at the interaction vertices attached to either atom. Dashed arrows represent photons of the external field which are annihilated at either atom. Time runs along the vertical axis from 0 up to the observation time $t$. A horizontal dash line indicates the state of the system at the observation time.} \label{fig2}
	\end{center}
\end{figure}

 \noindent Scattering involves processes in which  the atomic state  in  $|\Psi_{n}^{f}\rangle$ coincides with that in  $|\Psi_{n}(0)\rangle$, 
 but one of the probe photons in $|\Psi_{n}(0)\rangle$ is replaced in 
 $|\Psi_{n}^{f}\rangle$ with a scattered photon of undefined frequency $\omega'$, momentum $\mathbf{k}'$, and polarization $\bm{\epsilon}'$ upon integration. Depending on whether the active atom is in the steady state $g$ or $e$, the initial and final states read in each process,
\begin{align}
|\Psi_{1,3,6}(0)\rangle&=|\Psi_{0}\rangle_{g},\:\:|\Psi_{1,3,6}^{f}\rangle=\sum_{\mathbf{k}',\bm{\epsilon}'}|(N-1)_{\mathbf{k},\bm{\epsilon}},1_{\mathbf{k}',\bm{\epsilon}'};g,g\rangle,\nonumber\\
|\Psi_{2,4,5,7}(0)\rangle&=|\Psi_{0}\rangle_{e},\:\:|\Psi_{2,4,5,7}^{f}\rangle=\sum_{\mathbf{k}',\bm{\epsilon}'}|(N-1)_{\mathbf{k},\bm{\epsilon}},1_{\mathbf{k}',\bm{\epsilon}'};e,g\rangle.\nonumber
\end{align}
The collective scattering comprises two kinds of processes, namely, those in which the emitted photon is that which is exchanged between the atoms --see diagrams (1,2) of Fig.\ref{fig2}, and those in which the emitted photon appears in the diagrams as created and annihilated at the same atom, either $A$, diagram (3); or $B$, diagrams (4-7) and their hermitian conjugates (H.c.). The former are the result of the interference between the amplitudes of single scattering from each atom --see Fig.3 of Ref.\cite{myPRA}.

Same as for the case of single atom scattering,  under continuous illumination, $\gamma t\gg1$, the alternate processes of absorption and emission of photons take place without delay one after the other. This makes the intermediate atomic states to be transient and scattering to be continuous, from which follows that the scattered power is the time derivative of the EM energy in the final state of the corresponding diagrams, 
\begin{align}
\mathcal{W}^{coll}_{sc}&=\sum_{n=1}^{7}\frac{\mathop{d}}{\mathop{dt}}\langle\Psi_{n}(t)|\Psi_{n}^{f}\rangle\langle\Psi_{n}^{f}|H_{EM}|\Psi_{n}^{f}\rangle
\langle\Psi_{n}^{f}|\Psi_{n}(t)\rangle,\nonumber\\
&\gamma t\gtrsim1, \label{powerscat} 
\end{align}
where the script $coll$ stands for \emph{collective}. The contribution of the processes (1,2) is proportional to $\boldsymbol{\mu}\cdot \text{Im}\mathbb{G}(\omega ;\textbf{R})\cdot\boldsymbol{\mu}$, which is the result of the interference between the amplitudes of single-atom scattering. The contribution of the reminding processes, (3-7), is proportional to $\gamma_{\omega}$ instead, and the dependence on $\boldsymbol{\mu}\cdot\mathbb{G}(\omega ;\textbf{R})\cdot\boldsymbol{\mu}$ in that case results from the correlation induced by the interatomic interaction. 
%of all the seven different processes (and H.c.) contain terms of the form
%\begin{align}\label{scat2}
%\mathcal{W}^{col}_{sc}
%	\sim \frac{\mathcal{P},\gamma}{\Gamma}\;\frac{-\Omega_{0}^{2} \omega_{0}}{\epsilon_{0} c^{2}} \frac{\gamma \Gamma \omega^{2} \boldsymbol{\mu}\cdot \text{Im}\mathbb{G}(\omega ;\textbf{R})\cdot\boldsymbol{\mu}}{[(\omega-\omega_{0})^{2}+\frac{\Gamma^{2}}{4}](\omega-\omega_{0})^{2}+\frac{\gamma^{2}}{4}]}.
%\end{align}
%In this equation the prefactor $(\mathcal{P},\gamma)/\Gamma$ is the population of the states $g,e$ in the active atom. As for the rest of the factors, it is intended that $\omega$ and $\omega_{0}$ are of the same order of magnitude, and so are $\Gamma$, $\gamma$, $\gamma_{0}$ and $(\omega-\omega_{0})$. In particular, the above equation with prefactor $\mathcal{P}/\Gamma$ corresponds to the scattering power associated to the diagram (2) of Fig.\ref{xxxx}. Its calculation is given in full detail in Eq.(\ref{scat2expr}) of the Appendix \ref{app1} together with that of diagram (4). The dependence proportional to $\boldsymbol{\mu}\cdot \text{Im}\mathbb{G}(\omega ;\textbf{R})\cdot\boldsymbol{\mu}$ is common to all the scattering processes. In the case of the diagrams (1) and (2) it is a result of the interference between the single atom scattering amplitudes. In the case of diagrams (3-7) it is the result of the correlation induced on the scattering of each atom by effects of the EM interaction between them. 
Hereafter, we will find convenient to define 
\begin{align}
	\Omega(R)&=\frac{k_0^2}{\hbar\epsilon_{0}} \boldsymbol{\mu}\cdot \left[\textrm{Re}\mathbb{G}(\omega;\mathbf{R}) + i \textrm{Im}\mathbb{G}(\omega;\mathbf{R}) \right]\cdot\boldsymbol{\mu}\nonumber\\
	&\equiv \widetilde{\Omega}(R)-i\widetilde{\Gamma}(R).
\end{align}
This frequency is indeed the time derivative of the time evolution operator element from $|e,g\rangle$ to $|g,e\rangle$ at leading order in $W$, i.e., the complex-valued rate of the excitation transfer \cite{PRA3_Julio}. The calculation of the power scattered associated to diagrams (2) and (4) is given in detail in Appendix \ref{app1}.

As for the collective scattering cross-section it can be read off from Eq.(\ref{powerscat}), normalizing the scattered power by the intensity of the  probe field, i.e.,
\begin{equation}
		\sigma^{coll}_{sc}= \frac{2\hbar\omega}{c\epsilon_{0}E_{0}^{2}}\sum_{i=1}^{7}\frac{\mathop{d P_{i}(t)}}{\mathop{dt}},\quad
		\gamma t\gtrsim1.
\end{equation}
Putting together the contributions of the single atom  ($sing$) and collective scattering, $\sigma_{sc}=\sigma^{sing}_{sc}+\sigma^{coll}_{sc}$, the complete expression is
\begin{widetext}
\begin{equation}\label{scatty}
\begin{split}
\frac{\sigma_{sc}}{\sigma_{0}}
& =  \frac{\gamma_{0}^{2}}{4[(\omega-\omega_{0})^{2}+\frac{\Gamma^{2}}{4}]} +  \frac{\gamma_{0}^{2}}{4[(\omega-\omega_{0})^{2}+\frac{\gamma^{2}}{4}]}+ \frac{\mathcal{P}}{\Gamma}\; \left[ \frac{\left[[(\omega-\omega_{0})^{2}-\frac{\gamma \Gamma}{4}]\widetilde{\Gamma}(R)-(\omega-\omega_{0})(\frac{\gamma+\Gamma}{2})\widetilde{\Omega}(R)\right]\gamma_{0}^{2}}{(\gamma+\Gamma)[(\omega-\omega_{0})^{2}+\frac{\Gamma^{2}}{4}][(\omega-\omega_{0})^{2}+\frac{\gamma^{2}}{4}]} \right. \\
&  -  \left. \frac{\widetilde{\Gamma}(R) \gamma_{0}\left[(\omega-\omega_{0})^{2}-\frac{\gamma \Gamma}{4} -\frac{\gamma_{0}\Gamma}{4}\right]}{[(\omega-\omega_{0})^{2}+\frac{\Gamma^{2}}{4}][(\omega-\omega_{0})^{2}+\frac{\gamma^{2}}{4}]} -\frac{ \widetilde{\Gamma}(R) \gamma_{0}^{2}}{\left( \gamma+\Gamma \right)[(\omega-\omega_{0})^{2}+\frac{\gamma^{2}}{4}]} \right] \\
& +\frac{\gamma}{\Gamma}\; \left[ \frac{\left[(\omega-\omega_{0})^{2}+\frac{\gamma \Gamma}{4}\right]\widetilde{\Gamma}(R) \gamma_{0} +(\omega-\omega_{0})\widetilde{\Omega}(R)\gamma_{0}^{2}-\frac{\gamma+\Gamma}{4}\widetilde{\Gamma}(R)\gamma_{0}^{2}}{[(\omega-\omega_{0})^{2}+\frac{\Gamma^{2}}{4}][(\omega-\omega_{0})^{2}+\frac{\gamma^{2}}{4}]} \right],
\end{split}
\end{equation}
\end{widetext}
where $\sigma_{0}=\frac{2 \omega_{0} \mu_{||}^{2}}{\epsilon_{0} c \hbar \gamma_{0}}$ is the scattering cross-section of a single atom in free space, with $\mu_{\parallel}=\boldsymbol{\mu}\cdot\bm{\epsilon}$ the transition dipole moment along the polarization vector of the probe field.

%\begin{align}
%\sigma^{col}_{sc}&= \frac{2\hbar\omega}{c\epsilon_{0}E_{0}^{2}}\sum_{i=1}^{7}\frac{\mathop{d P_{i}(t)}}{\mathop{dt}},\:\Gamma t\gtrsim1,\\
%&\sim \frac{\mathcal{P},\gamma}{\Gamma}\;\sigma_{0}\gamma_{0} \frac{-\omega^{2} \boldsymbol{\mu}\cdot \text{Im}\mathbb{G}(\omega %;\textbf{R})\cdot\boldsymbol{\mu}}{[(\omega-\omega_{0})^{2}+\frac{\Gamma^{2}}{4}](\omega-\omega_{0})^{2}+\frac{\gamma^{2}}{4}]}, \label{sigmascat} 
%\end{align}

\subsection{Collective absorption and stimulated emission}
\begin{figure}
	\begin{center}
		\includegraphics[width=82mm,angle=0,clip]{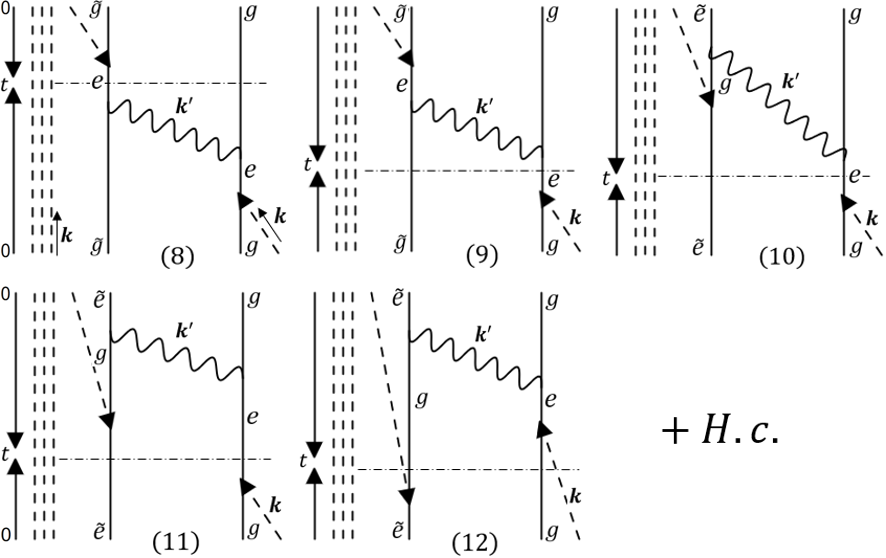}
		\caption{Diagrammatic representation of the collective absorption processes.} \label{fig3}
	\end{center}
\end{figure}

\noindent Generically, absorption processes are those in which the initial and final states, $|\Psi_{n}(0)\rangle$ 
and $|\Psi_{n}^{f}\rangle$, differ both in the atomic internal states and in the number of probe field photons. Here we concentrate in those processes in which the atoms gets correlated through the exchanged of a virtual photons when either absorbing or emitting a probe field photon. They are 
 represented by the diagrams (8-12) of Fig.\ref{fig3} together with their hermitian conjugates. We can distinguish processes in which either the active or the passive atom absorbs a probe field photon and transits from the ground state $g$ to the upper level $e$ [diagrams (8-11)], that we identify with collective absorption; and one process in which the active atom de-excites at the time it emits a photon of the same frequency and polarization as those of the probe field [diagram (12)], that we identify with stimulated emission. Hence, for the initial and final states of collective absorption  we have
 \begin{align}
 	|\Psi_{8}(0)\rangle&=|\Psi_{0}\rangle_{g},\:\:|\Psi_{8}^{f}\rangle=|(N-1)_{\mathbf{k},\bm{\epsilon}};e,g\rangle,\nonumber\\
 	|\Psi_{9}(0)\rangle&=|\Psi_{0}\rangle_{g},\:\:|\Psi_{9}^{f}\rangle=|(N-1)_{\mathbf{k},\bm{\epsilon}};g,e\rangle.\nonumber\\
 	|\Psi_{10,11}(0)\rangle&=|\Psi_{0}\rangle_{e},\:\:|\Psi_{10,11}^{f}\rangle=|(N-1)_{\mathbf{k},\bm{\epsilon}};e,e\rangle,\nonumber
 \end{align}
 while for stimulated emission,
\begin{equation}
	|\Psi_{12}(0)\rangle=|\Psi_{0}\rangle_{e},\quad|\Psi_{12}^{f}\rangle=|(N+1)_{\mathbf{k},\bm{\epsilon}};g,g\rangle.\nonumber
\end{equation}
 In contrast to the processes contributing to scattering, absorption and emission of photons in these processes do not take place in a continuous manner, but just at an average  rate. Under steady atomic populations and continuous  illumination,  the transitions  induced by the potential $W$ 
 from the steady states of the active atom, $\tilde{g}\rightarrow e$, $\tilde{e}\leftrightarrow g$,  take place at a constant rate $\Gamma$, as that is the coherence time sets by the pump. Likewise, the transitions $e\leftrightarrow g$ between the states of the passive atom take place at a rate $\gamma$, as it is not subjective to the pump. Thus, the contribution to the absorption rate of a process in which  the atom that undergoes the excitation is the active one, is $\Gamma$ times the probability of excitation for asymptotic times $\Gamma t\gg1$; and $\gamma$ times the probability of excitation for asymptotic times $\gamma t\gg1$ when the atom to be excited is the passive one. On the other hand, since the states $|\Psi_{0}\rangle_{g}$ and  $|\Psi_{0}\rangle_{e}$ are stationary, absorption is necessarily  
 followed  by processes that take the  
 atomic state at time $t$ back to the stationary state at time $0$. In particular, the state $e$ decays  into $g$  
 either emitting a scattered photon of frequency $\omega$ 
  or by nonraditive means. Therefore, in order not to double-count the probability of radiative decay, the contribution of 
 the processes associated to  scattering must be substracted from that of the absorption ones [diagrams (8-11)], resulting in nonradiative absorption only. In particular the scattering processes to be substracted from each absorption process are as follows, from diagram (8), diagram (3); from diagram (9), diagram (6); from diagram (10), diagram (5) \cite{foot1}; from diagram (11), diagram (4). Likewise, the state $g$ of the active atom at the observation time in diagram (12)   
 ends up transiting to the state $\tilde{e}$ under the action of the pump, which is only accompanied by the spontaneous emission of 
 photons of frequency $\simeq\omega_{u}-\omega_{0}$. Since no other radiative processes are involved there, %and radiation of frequency $\omega_{u}-\omega_{0}$ is irrelevant to us, 
 no double-counting is associated to the probability of stimulated emission, and no additional substraction is necessary. Thus, the power absorbed by the system is written as
\begin{align}
		\mathcal{W}^{coll}_{abs}
		& =  \Gamma \sum_{n=8,12} \Bigl[\bra{\Psi_{n}(0)}H_{EM}\ket{\Psi_{n}(0)}\nonumber\\
		&-\langle\Psi_{n}(t)|\Psi_{n}^{f}\rangle\bra{\Psi_{n}^{f}}H_{EM}\ket{\Psi_{n}^{f}} \langle\Psi_{n}^{f}|\Psi_{n}(t)\rangle \Bigr]\nonumber \\
		& + \gamma \sum_{n=9}^{11}\Bigl[\bra{\Psi_{n}(0)}H_{EM}\ket{\Psi_{n}(0)}\nonumber\\
		&-\langle\Psi_{n}(t)|\Psi_{n}^{f}\rangle \bra{\Psi_{n}^{f}}H_{EM}\ket{\Psi_{n}^{f}} \langle\Psi_{n}^{f}|\Psi_{n}(t)\rangle\Bigr]\nonumber \\
		& - \sum_{n=3}^{6}\frac{d}{dt} \langle\Psi_{n}^{f}|\Psi_{n}(t)\rangle \bra{\Psi_{n}^{f}}H_{EM}\ket{\Psi_{n}^{f}} \langle\Psi_{n}^{f}|\Psi_{n}(t)\rangle,\nonumber\\
		&\qquad\gamma t\gtrsim1.\label{Wcollabs}
\end{align}
The calculations of the absorbed power and emitted power associated to the processes (9) and (12), respectively, are given in full detail in  Appendix \ref{app1}.
 
 Again, as for the collective cross-section of absorption, it can be read off from  Eq.(\ref{Wcollabs}), normalizing the absorbed power by the intensity of the  probe field, i.e.,
 \begin{align}
 	\sigma^{coll}_{abs}&= \frac{2\hbar\omega}{c\epsilon_{0}E_{0}^{2}}\Bigl[\Gamma\Bigl(P_{8}(t)-P_{12}(t)\Bigr)+\gamma\Bigl(P_{9}(t)+P_{10}(t)\nonumber\\
 	&+P_{11}(t)\Bigr)-\sum_{n=3}^{6}\frac{\mathop{d P_{i}(t)}}{\mathop{dt}}\Bigr],\quad
 	\gamma t\gtrsim1.
 \end{align}
Adding up the contributions of single-atom and collective absorption, $\sigma_{abs}=\sigma^{sing}_{abs}+\sigma^{coll}_{abs}$, the total absorption cross-section reads
\begin{widetext}
\begin{equation}
	\begin{split}
		\frac{\sigma_{abs}}{\sigma_{0}}
		& =  \frac{\gamma_{0}\left(\frac{\gamma}{\Gamma}\gamma_{nr}-\frac{\mathcal{P}}{\Gamma}\mathcal{P}\right)}{4[(\omega-\omega_{0})^{2}+\frac{\Gamma^{2}}{4}]}  + \frac{\gamma_{0}\gamma_{nr}}{4[(\omega-\omega_{0})^{2}+\frac{\gamma^{2}}{4}]} -		\frac{\mathcal{P}}{\Gamma}\left[\frac{  \tilde{\Gamma}(R)\gamma_{0}\gamma_{nr}}{(\gamma+\Gamma)[(\omega-\omega_{0})^{2}+\frac{\gamma^{2}}{4}]} +\frac{  \tilde{\Gamma}(R)\gamma_{0}\Gamma}{(\gamma+\Gamma)[(\omega-\omega_{0})^{2}+\frac{\Gamma^{2}}{4}]}\right]\\
		&+  \frac{\gamma}{\Gamma}\; \left[\frac{[(\omega - \omega_{0})\tilde{\Omega}(R)- \frac{\Gamma}{2}\tilde{\Gamma}(R)]\gamma_{0}\gamma_{nr}}{2[(\omega-\omega_{0})^{2}+\frac{\Gamma^{2}}{4}][(\omega-\omega_{0})^{2}+\frac{\gamma^{2}}{4}]}+  \;\frac{[(\omega - \omega_{0})\tilde{\Omega}(R)- \frac{\gamma}{2}\tilde{\Gamma}(R)](\mathcal{P}+\gamma_{nr})\gamma_{0}}{2[(\omega-\omega_{0})^{2}+\frac{\Gamma^{2}}{4}][(\omega-\omega_{0})^{2}+\frac{\gamma^{2}}{4}]}\right].
	\end{split}
\end{equation}
\end{widetext}
In this equation, the first two terms correspond to single-atom absorption from the active and passive atom, respectively; the third and fourth terms correspond to  collective absorption from the stationary state $|\Psi_{0}\rangle_{g}$; the fifth term comes from collective absorption from the stationary state $|\Psi_{0}\rangle_{e}$; and the last two terms are those of single-atom and collective stimulated emission, respectively. 
\begin{figure}
	\begin{center}
		\includegraphics[height=72mm,width=80mm,angle=0,clip]{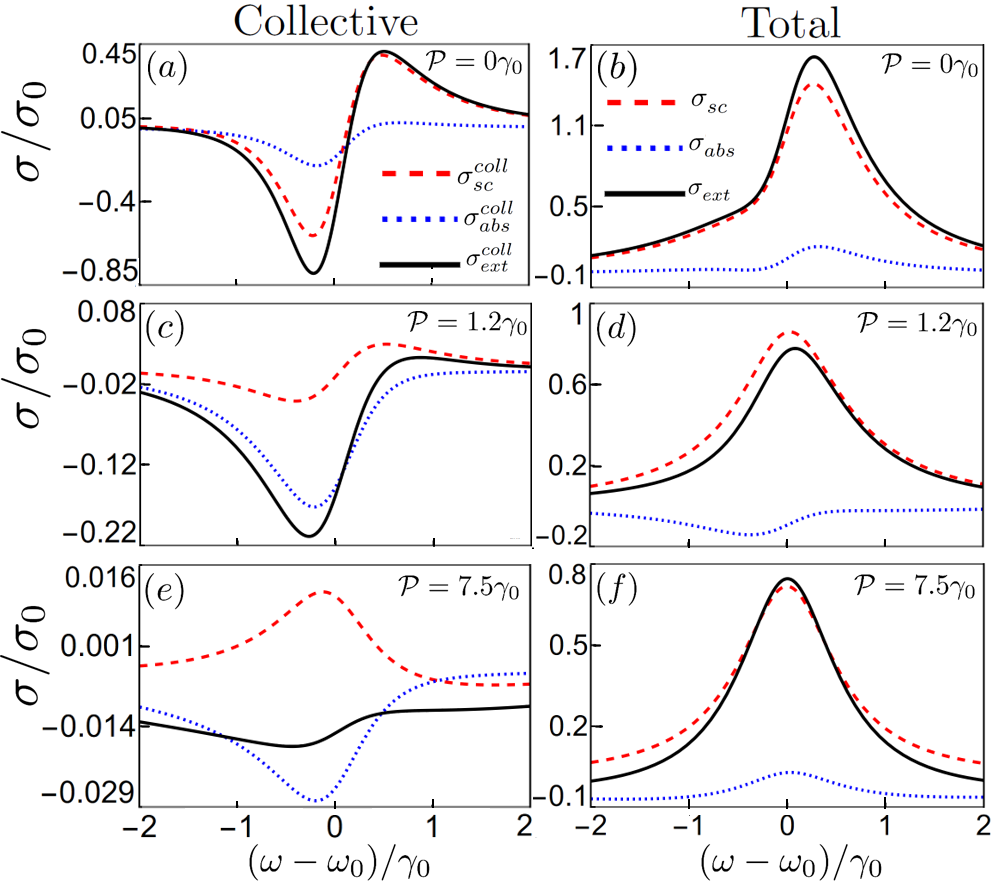}
		\caption{Graphical representation of the collective (left panels) and total (right panels) cross-sections as a function of the detuning $(\omega-\omega_{0})$, for fixed interatomic distance $k_{0}R\simeq2$, fixed non-radiative decay rate $\gamma_{nr}=\gamma_{0}/5$, and different values of the pump rate. That is, 
			  $\mathcal{P}=0$ [(a)-(b)],  $\mathcal{P}=1.2\gamma_{0}$ [(c)-(d)],  $\mathcal{P}=7.5\gamma_{0}$ [(e)-(f)].} \label{fig11}
	\end{center}
\end{figure}

The collective extinction cross-section $\sigma^{coll}_{ext}$ results from the addition of $\sigma^{coll}_{sc}$ and $\sigma^{coll}_{abs}$. In Fig.\ref{fig11} we represent graphically the collective (left panels) and total (right panels) cross-sections of the binary system, as a function of the laser detuning $(\omega-\omega_{0})$, for fixed interatomic distance $k_{0}R\simeq2$, fixed non-radiative decay rate $\gamma_{nr}=0.2\gamma_{0}$, and different values of the pump rate. In the first place, we observe a strong asymmetry of the collective scattering cross-section with respect to the laser detuning.  Second, while the increase of pump rate  enhances the collective scattering cross-section at resonance, it reduces total scattering. In addition, the increase of the pump rate undermines the collective response generally.

\begin{figure}
	\begin{center}
		\includegraphics[height=30mm,width=80mm,angle=0,clip]{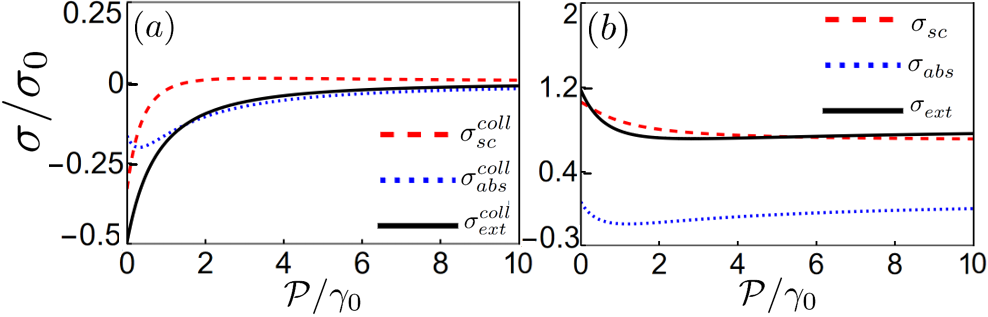}
		\caption{Graphical representation of the collective [left panel (a)] and total [right panel (b)] cross-sections as a function of the pump rate $\mathcal{P}$, at exact resonance, $\omega=\omega_{0}$, fixed interatomic distance $k_{0}R\simeq2$, and fixed non-radiative decay rate $\gamma_{nr}=\gamma_{0}/5$.} \label{fig12}
	\end{center}
\end{figure}

The latter is confirmed by the graphs of Fig.\ref{fig12}, in which the collective and total cross-sections are represented against the pump rate, at resonance, $\omega=\omega_{0}$, and fixed interatomic distance $k_{0}R\simeq2$. That is, the collective cross-sections vanish asymptotically for high pump rates. In contrast, the total extinction cross-section decreases with the pump rate and tends to $3\sigma_{0}/4$ asymptotically. 

\begin{figure}
	\begin{center}
		\includegraphics[height=72mm,width=80mm,angle=0,clip]{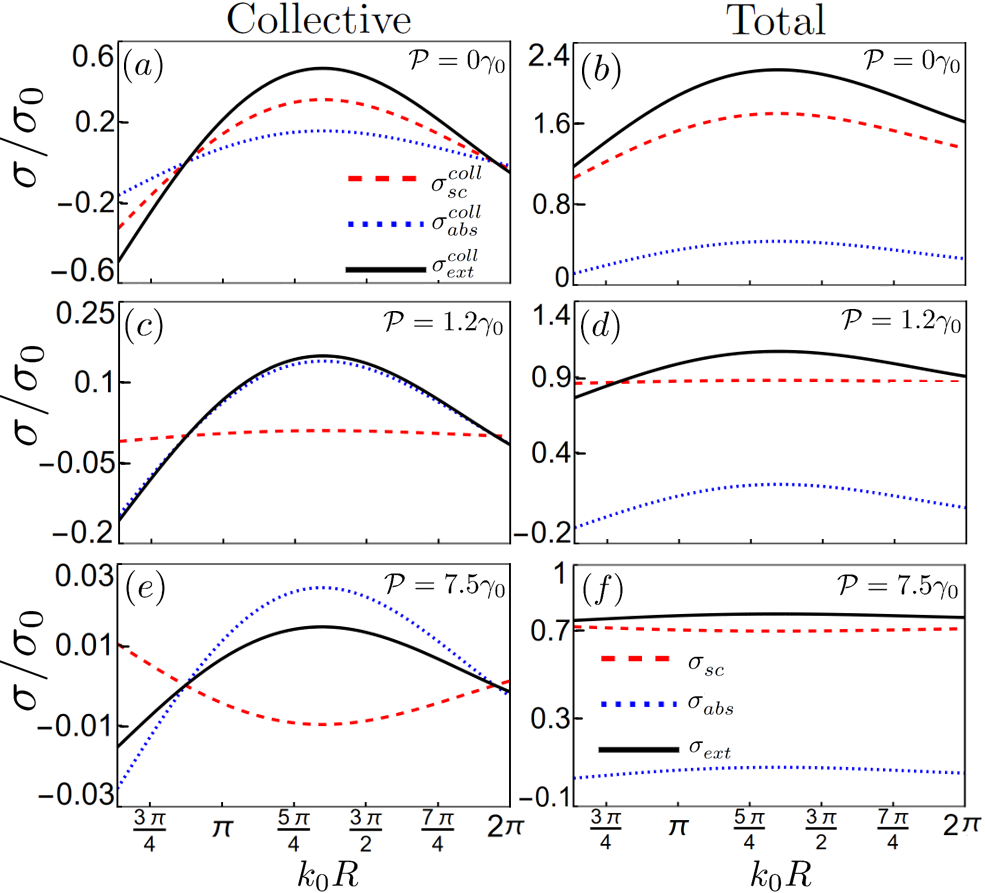}
		\caption{Graphical representation of the collective (left panels) and total (right panels) cross-sections as a function of the detuning interatomic distance, $k_{0}R$, at exact resonance, $\omega=\omega_{0}$,  fixed non-radiative decay rate $\gamma_{nr}=\gamma_{0}/5$, and different values of the pump rate. That is, 
			$\mathcal{P}=0$ [(a)-(b)],  $\mathcal{P}=1.2\gamma_{0}$ [(c)-(d)],  $\mathcal{P}=7.5\gamma_{0}$ [(e)-(f)].} \label{fig13}
	\end{center}
\end{figure}

Finally, the cross-sections are represented in Fig.\ref{fig13} as a function of the interatomic distance. It is apparent that the collective effects become more relevant at shorter distances.  In this respect, we remind that the perturbative nature of our approach obliges us to impose a cutoff on the interatomic distance in all our calculations. The cut-off $k_{0}R\gtrsim2$ is chosen to guarantee that $|\tilde{\Omega}(R)|,|\tilde{\Gamma}(R)|\lesssim\gamma_{0}$, which allows us to truncate our calculation at leading order in the interaction potential $W$. In turn, this is in accordance with the fact that the total absorption cross-section must be positive definite at zero pump.

\subsection{Collective spontaneous emission}
 
\begin{figure}
 	\begin{center}
 		\includegraphics[width=82mm,angle=0,clip]{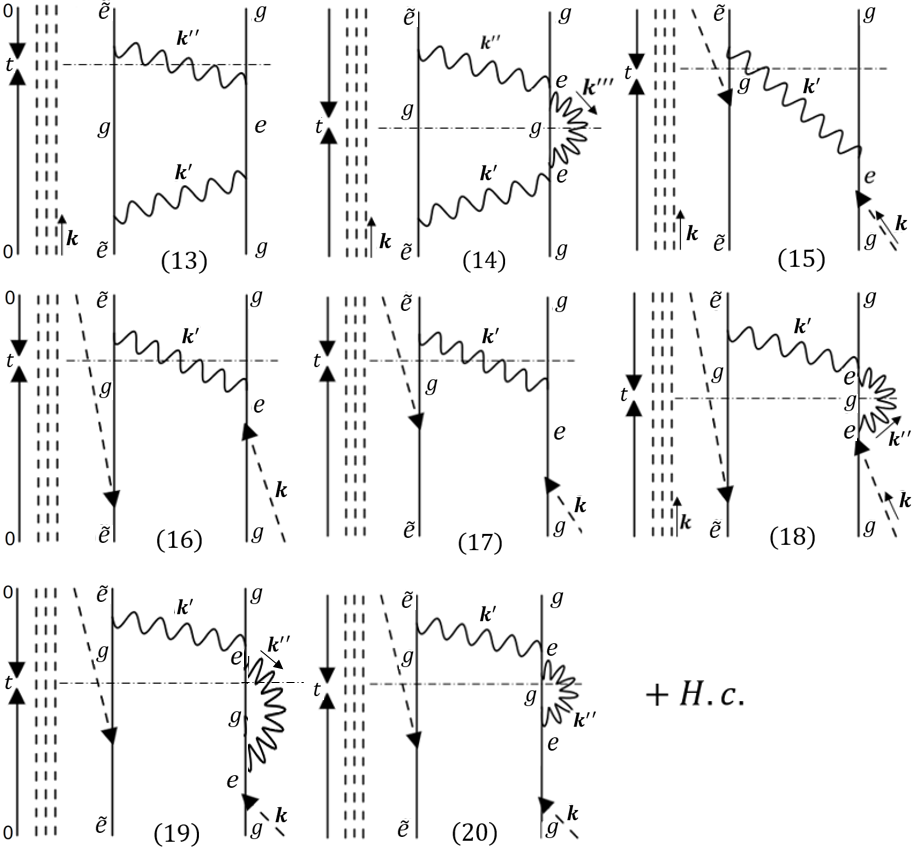}
 		\caption{Diagrammatic representation of the collective spontaneous emission processes.} \label{fig4}
 	\end{center}
 \end{figure}

Collective spontaneous emission, at leading order, corresponds to the processes depicted by diagrams 
 (13-20) in Fig.\ref{fig4} and their hermitian conjugates. In all of them, the active atom transits from the excited state at the initial time $0$ to the ground state at the observation time $t$, whereas the passive atom either remains in its ground state $g$ [diagrams (13,15-17)] or decays into $g$ after having absorbed a resonant photon emitted by the active atom [diagrams (14,18-20)]. Thus, the final states $|\Psi_{n}^{f}\rangle$, $n=13,..,20$,   
 contain the same number of probe field photons as the initial states in each case, 
 $|\Psi_{n}(0)\rangle$, plus one more photon of undefined frequency, momentum and polarization upon integration. That is,
 \begin{equation}
 	|\Psi_{13-20}(0)\rangle=|\Psi_{0}\rangle_{e},\:\:|\Psi_{13-20}^{f}\rangle=\sum_{\mathbf{k}',\bm{\epsilon}'}|N_{\mathbf{k},\bm{\epsilon}},1_{\mathbf{k}',\bm{\epsilon}'};g,g\rangle.\nonumber
 \end{equation} 

 Further, amongst these processes we can distinguish those in which spontaneous emission is independent of the probe field, diagrams (13) and (14), from those in which the emission is induced by the coupling to probe field photons, diagrams (15-20). The latter, of higher order, are proportional to the intensity of the probe field, carrying a prefactor of the order of $\Omega_{0}^{2}/\gamma^{2}$ or $\Omega_{0}^{2}/\Gamma^{2}$.  Same as in the case of collective absorption,  under steady conditions, the rate at which the system undergoes 
 spontaneous emission in each process is  $\gamma$ or $(\Gamma+\gamma)/2$ times the probability of the process for asymptotic times, $\gamma t\gg1$. The choice of the prefactor  $\gamma$ or $(\Gamma+\gamma)/2$ depends in each case on whether the emitted photon flies between both atoms or it appears as created and annihilated at the passive atom, respectively. Spontaneous emission 
 is inherently incoherent (\emph{inc}), and the incoherent power associated to the collective spontaneous emission reads $\mathcal{W}_{inc}^{coll}=\hbar\omega\Gamma_{0}^{coll}$, where $\Gamma_{0}^{coll}$ is given by   
 \begin{align}
 	\Gamma^{coll}_{0}&=\frac{\Gamma+\gamma}{2}\sum_{n=13,15}^{17}2\textrm{Re}\Bigl[\langle\Psi_{n}(t)|\Psi_{n}^{f}\rangle\langle\Psi_{n}^{f}|H_{EM}|\Psi_{n}^{f}\rangle
 	\nonumber\\
 	&\times\langle\Psi_{n}^{f}|\Psi_{n}(t)\rangle
 	-\langle\Psi_{n}(0)|H_{EM}|\Psi_{n}(0)\rangle\Bigr]\nonumber\\
 	&+\gamma\sum_{n=14,18}^{20}2\textrm{Re}\Bigl[\langle\Psi_{n}(t)|\Psi_{n}^{f}\rangle
 	\langle\Psi_{n}^{f}|H_{EM}|\Psi_{n}^{f}\rangle\nonumber\\
 	&\times\langle\Psi_{n}^{f}|\Psi_{n}(t)\rangle
 	-\langle\Psi_{n}(0)|H_{EM}|\Psi_{n}(0)\rangle\Bigr],\quad\gamma t\gtrsim1.\nonumber
 \end{align}
 A calculation analogous to that of the previous sections yields, for the total spontaneous emission rate $\Gamma_{0}$,
 
% \textcolor{red}{LORENA TIENE QUE CORREGIR LA FORMULA CON LOS PREFACTORES ATENUANTES CORRECTOS, $(\Gamma+\gamma)/2$ Y $\gamma$ EN LA FORMULA (60) DEL DRAFT, EPIGRAFE 4.9}
 
 \begin{widetext}
 \begin{equation}
 \begin{split}
 \Gamma_{0}
 & =\frac{\mathcal{P}}{\Gamma} \;  \left\{ \gamma_{0}\left[1-\frac{\Omega_{0}^{2}\Gamma^{2}/16}{[(\omega-\omega_{0})^{2}+\Gamma^{2}/4]^{2}}\right]-\left(\frac{\gamma+\Gamma}{2}\right)\frac{8\: \tilde{\Gamma}^{2}(R)} {\Gamma (\gamma+\Gamma)} + \; \gamma\frac{8[\tilde{\Omega}^{2}(R)+\tilde{\Gamma}^{2}(R)]\gamma_{0}}{\Gamma (\gamma+\Gamma)^{2}} \right. \\
 & +  \frac{\Omega_{0}^{2}}{4\hbar^2} \left\{ \left(\frac{\gamma+\Gamma}{2}\right) \left[\frac{2(\omega-\omega_{0})\tilde{\Omega}(R) \; +4 \tilde{\Gamma}(R) \left[\frac{(\omega-\omega_{0})^{2}}{\Gamma}+\frac{\gamma}{4}\right] -\gamma \tilde{\Gamma}(R)}{\left[(\omega-\omega_{0})^{2}+\frac{\gamma^{2}}{4}\right]\left[(\omega-\omega_{0})^{2}+\frac{\Gamma^{2}}{4}\right]}+  \; \frac{8 (\omega-\omega_{0}) \tilde{\Omega}(R)}{\Gamma (\gamma+\Gamma) \left[(\omega-\omega_{0})^{2}+\frac{\Gamma^{2}}{4}\right]} \right. \right.\\
 &- \left. \frac{8 (\omega-\omega_{0})\tilde{\Omega}(R)}{\Gamma (\gamma+\Gamma) \left[(\omega-\omega_{0})^{2}+\frac{\gamma^{2}}{4}\right]}\right]\; + \gamma\gamma_{0}\left[\frac{-8 [(\omega-\omega_{0})\tilde{\Omega}(R) - \frac{\Gamma}{2} \tilde{\Gamma}(R)]}{\Gamma (\gamma + \Gamma)^{2} \left[(\omega-\omega_{0})^{2}+\frac{\Gamma^{2}}{4}\right]}+ \; \frac{8 [(\omega-\omega_{0}) \tilde{\Omega}(R)+\frac{\gamma}{2} \tilde{\Gamma}(R)]}{\Gamma (\gamma+\Gamma)^{2} \left[(\omega-\omega_{0})^{2}+\frac{\gamma^{2}}{4}\right]} \right. \\
 & -\left. \left.\left.  \; \frac{2\left[(\omega-\omega_{0})\tilde{\Omega}(R)(1-\gamma/\Gamma)+2\tilde{\Gamma}(R)\left((\omega-\omega_{0})^{2}/\Gamma+\gamma/4\right)\right]}{(\gamma+\Gamma) \left[(\omega-\omega_{0})^{2}+\frac{\gamma^{2}}{4}\right] \left[(\omega-\omega_{0})^{2}+\frac{\Gamma^{2}}{4}\right]}\right]  \right\} \right\}.
 \end{split}
 \end{equation}
\end{widetext}
 In this equation, the first two terms correspond to single-atom decay \cite{myPRA}, of orders unity and $\Omega^{2}_{0}/\Gamma^2$. As for the rest, the collective decay contain laser-independent terms of order $|\Omega(R)|^{2}/\Gamma^{2}$ and laser-dependent terms of order $|\Omega(R)|\Omega^{2}_{0}/\Gamma^3$.
 
 % \textcolor{red}{EN LAS NOTAS SE DICE QUE AL AUMENTAR EL PUMP FIELD, EL RESULTADO TIENDE A gamma_0, pero eso no significa nada más que que el átomo excitado decae a rate constante como si estuviese en espacio libre}

\subsection{Resonant energy transfer}\label{subplot}

\begin{figure}
	\begin{center}
		\includegraphics[width=82mm,angle=0,clip]{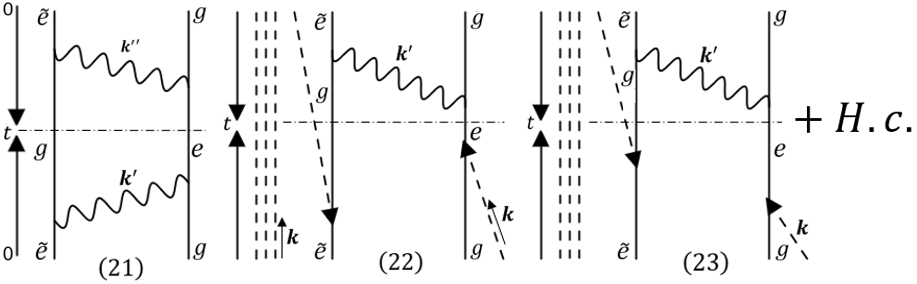}
		\caption{Diagrammatic representation of the  processes contributing to resonant energy transfer.} \label{fig5}
	\end{center}
\end{figure}

Besides the aforementioned radiative processes, there exists an additional photon-mediated phenomenon which does not involve radiation and is readily identifiable with resonant energy transfer (RET). Its diagrammatic representation is given in Fig.\ref{fig5}. In all the diagrams there, the excitation of the active atom is transferred to the passive atom without emission or absorption of radiation, but mediated in most of the cases by virtual photons  [diagrams (22-24)]. Thus, the initial and final states are, in all the cases,
 \begin{equation}
	|\Psi_{21-23}(0)\rangle=|\Psi_{0}\rangle_{e},\:\:|\Psi_{21-23}^{f}\rangle=|N_{\mathbf{k},\bm{\epsilon}};g,e\rangle.\nonumber
\end{equation} 
In all these processes, the final state is made of two unstable states, namely, $g$ for the active atom subjected to the pump, and $e$ for the passive atom subjective to free spontaneous emission. As a result,  under steady conditions, the rate of resonant energy transfer $\Gamma_{RET}$ is  $(\Gamma+\gamma)/2$ times the addition of the probability of transfer associated to all the processes for asymptotic times, $\gamma t\gg1$,
\begin{widetext}
\begin{equation}
\begin{split}
\Gamma_{RET}
&=\frac{\Gamma+\gamma}{2}\sum_{n=21}^{23}|\langle\Psi_{n}(t)|\Psi_{n}^{f}\rangle|^{2}\\
& = \left(\frac{\Gamma+\gamma}{2}\right) \; \frac{\mathcal{P}}{\Gamma} \; \left[8\frac{\tilde{\Omega}^{2}(R) +\tilde{\Gamma}^{2}(R)}{(\gamma + \Gamma)^{2}} \right.
 + \left. \frac{2\Omega_{0}^{2}}{\hbar^{2}} \; \left(\frac{[(\omega-\omega_{0})\; \tilde{\Omega}(R)+\frac{\gamma}{2} \; \tilde{\Gamma}(R)]}{(\gamma+\Gamma)^{2} [(\omega-\omega_{0})^{2}+\frac{\gamma^{2}}{4}]}
- \;\frac{[(\omega-\omega_{0})\; \tilde{\Omega}(R)-\frac{\Gamma}{2} \;  \tilde{\Gamma}(R)]}{(\gamma+\Gamma)^{2} [(\omega-\omega_{0})^{2}+\frac{\Gamma^{2}}{4}]}\right)\right].
\end{split}
\end{equation}
\end{widetext}
In this equation the first term corresponds to diagram (21), which is independent of the probe field. It is of order $\Gamma(|\Omega(R)|^{2}/\Gamma^{2})$. As for the rest of the diagrams, associated to probe-field-induce energy transfer, their contribution is of order  $\Gamma(|\Omega(R)|\Omega_{0}^{2}/\Gamma^{3})$, and thus, of higher order. 

\section{Discussion and Conclusions}\label{lasec4}
In this article we have characterized the optical response of a binary system of identical atoms in which one of them is incoherently excited. Following up the study in Ref.\cite{myPRA}, we have analysed the impact of the pump in the scattering and absorptive properties of the system, in relation with the dynamics of the internal atomic states. %Hence, this study is part of a series of works intended to characterize the optical response of atomic ensembles which are excited and arranged  in different manners, which is of application in the monitoring of the dynamics of atomic multi-qubits. %Besides, the effect of the pump in collective phenomena like resonant energy transfer have been analysed.

%Follow up the study in Ref.\cite{myPRA} for a single pumped atom, and is part of a series of works intended to characterize the optical response of atomic ensembles which are excited and arranged  in different manners. 

Here we have restricted ourselves to the weak coupling regime in considering the interatomic interaction and the interaction of the atoms with the probe field. That has allowed us to perform a diagrammatic identification of all the processes which contribute to scattering, absorption, stimulated emission, spontaneous emission and resonant energy transfer --Figs. \ref{fig2}, \ref{fig3}, \ref{fig4}, \ref{fig5}, respectively, and to carry out a perturbative computation of the associated rates, emission powers and cross-sections. Although the perturbative treatment of the interatomic interaction imposes a limit upon the interatomic distance, $k_{0}R\gtrsim2$, the generalization to shorter distances is not difficult to achieve. Hence, a non-perturbative treatment results in  effective renormalizations of the resonant frequency (analogous to the Autler–Townes splitting \cite{Scully,Autler}) and the linewidth  in all the Lorentzian formulas obtained in the perturbative approximation \cite{PRA4_Julio}. 

From a fundamental perspective, the most important contribution of this study is the identification and characterization of the nature of each of the processes of the optical response based on the physical content of the initial and final state of each one. Furthermore, for the quantum computation of the probability and power associated to each process, we have derived effective rules which account for the incoherent effects associated to spontaneous decay and incoherent pump. 

Phenomenologically, from a qualitative perspective we have analysed the interference effects induced by the interatomic distance in the emitted radiation, and the effects of the pump rate which determines the population of the atomic levels. From a quantitative perspective, according to the graphs of Figs. \ref{fig11},  \ref{fig12} and  \ref{fig13},    it is found that  for sufficiently strong pump the collective component of the extinction cross-section becomes negligibly small,  regardless of the interatomic distance, as gains compensate for losses, and the total extinction cross-section is reduced to less than half of its value in the absence of pumping.  In contrast, at weak pumping and short interatomic
distances, interference effects lead to a significant suppression of the extinction cross-section relative to that of two isolated atoms.

Finally, it is worth comparing our quantum approach with a semiclassical one, and our  results on a binary atomic system with those on a classical system of metallic dimers. 
Let us consider the system of Ref.\cite{ManjavacasPT1} consisting of a dimer of metallic spheres in which one of them is optically pumped. In such a case, the classical polarizability of the active element is formulated with a Lorentzian gain, which is intended to model an optical pump \cite{ManjavacasPT1,ChenetalPRA,ManjavacasSandersPT2}. %In the first place, only at zero pump both treatments agree. The reason for this is that the scattering processes reduce in that case to the ones depicted by diagrams (1), (3) and (6) of Fig.\ref{fig2}, which contain only one-photon intermediate states which yields mathematical expressions equivalent to those of the classical treatment using linear response theory. On the contrary, for non-zero pump, the results are quantitatively very different.
Several differences are found. In the first place, the classical calculation does not allow for the identification of spontaneous emission or stimulated emission, and neither of resonant energy transfer. %In addition, the quantum processes which involve excited initial  states contain two-photon intermediate processes, which makes them impossible to be described with classical linear response theory.
Quantitatively, for sufficiently strong pump, the classical approach allows for a negative extinction cross-section \cite{ManjavacasPT1}, which is in contradiction with the quantum result. In this regard, however, it is not clear to which extent the electric dipole approximation is sufficient to describe the interaction between the metallic spheres in the classical approach.

As for the generic differences between quantum and classical  approaches, in accordance with Ref.\cite{myPRA}, there is no satisfactory formulation for 
the linear polarizability of an excited atom that yields approximate results for its optical response. Hence, all the processes in which the initial state contains the excited state of the active atom include two-photon intermediate states, which makes unsuitable the used of semiclassical linear response theory. In addition, we note that the processes in which both atoms are initially in their ground state contain only one-photon intermediate states --see diagrams (1), (3) and (6) of Fig.\ref{fig2} for scattering. In such a case, the semiclassical approach based on the use of an atomic polarizability and linear response theory would seem appropriate. For the polarizability of a two-level atom in its ground state, $\alpha(\omega)$, one could take a Lorentzian model \cite{BermanBoydMilonni} or, equivalently, that derived in Ref.\cite{myPRA} under quasiresonant conditions. In either case, however, including the nonradiative effects in the linewidth, in the absence of pumping, $\Gamma=\gamma=\gamma_{0}+\gamma_{nr}$, one finds out an agreement between the cross-sections derived from the quantum and from the semiclassical approach, except for the fact that the semiclassical result fails to distinguish scattering from absorption. We interpret this as a generic failure of the semiclassical approach. That is, even for the case that semiclassical linear response theory is suitable to describe one-photon quantum processes, the implementation of nonradiative effects is not correctly accounted for with the simple substitution $\gamma_{0}\rightarrow\gamma$ in the atomic polarizability --see Eq.(\ref{equclas}) in Appendix \ref{appn} and compare with Eq.(\ref{scatty}).

\acknowledgments
We gratefully acknowledge helpful discussions with Alejandro Manjavacas.  
L.A. acknowledges financial support from the Spanish Ministry of Science, Innovation and Universities (MICIU) through an FPU fellowship (ref. FPU2022/00817). J.S.-C. 
 acknowledges financial support from the Doctorate Program Funds No. UVa 2021 and Banco Santander. The work of L.A. and J.S.-C. was supported by  the project QCAYLE (NextGenerationEU funds, PRTRC17.I1).

\appendix

\section{Diagram reading and quantum expressions}
In this Appendix we compile the  mathematical expressions of the power associated to some of diagrams which appear in the main text.

\begin{widetext}
\subsection{Collectively scattered, absorbed, and incoherently emitted powers}\label{app1}
The expression for the contribution of diagram (2) in Fig.\ref{fig2} to the collective scattering power, in which the emitted photons fly between both atoms, reads

\begin{equation}
	\begin{split}
		\mathcal{W}_{sc}^{(2)}&
		 %=\frac{\mathcal{P}}{\Gamma}
		 = \frac{d}{dt} \; 
		\langle\Psi_{2}(t)|\Psi_{2}^{f}\rangle\langle\Psi_{2}^{f}|H_{EM}|\Psi_{2}^{f}\rangle\langle\Psi_{2}^{f}|\Psi_{2}(t)\rangle\\
		& = \frac{\mathcal{P}}{\Gamma}\;\frac{d}{dt} 2\textrm{Re}\sum_{\textbf{k}', \boldsymbol{\epsilon}'}  \hbar \omega' \{ \hbar ^{-4}\int_{0}^{t} d\tau \int_{0}^{\tau} d \tau'\int_{0}^{t} d\tilde{\tau} \int_{0}^{\tilde{\tau}} d \tilde{\tau}' \bra{N_{\mathbf{\textbf{k}, \boldsymbol{\epsilon}}}; \;\tilde{e}, \;g} \mathbb{U}_{0}^{\dagger}(\tau')\ket{N_{\textbf{k}, \boldsymbol{\epsilon}}; \;\tilde{e}, \;g} \\
		& \times \; \bra{N_{\textbf{k}, \boldsymbol{\epsilon}}; \;\tilde{e}, \;g} \mathbf{d}\cdot\mathbf{E}_{\mathbf{\textbf{k}', \boldsymbol{\epsilon}'}}^{(+)}(\mathbf{r}_{A})\ket{N_{\textbf{k}, \boldsymbol{\epsilon}}, 1_{\textbf{k}'; \boldsymbol{\epsilon}'}; \;g, \;g}
		\bra{N_{\textbf{k}, \boldsymbol{\epsilon}}, 1_{\textbf{k}', \boldsymbol{\epsilon}'}; \;g, \;g} e^{-\frac{\Gamma}{2}(\tau-\tau')} \mathbb{U}_{0}^{\dagger}(\tau-\tau') \\		
		& \times \;\ket{N_{\textbf{k}, \boldsymbol{\epsilon}}, 1_{\textbf{k}', \boldsymbol{\epsilon}'}; \;g, \;g}\bra{N_{\textbf{k}, \boldsymbol{\epsilon}}, 1_{\textbf{k}', \boldsymbol{\epsilon}'}; \;g, \;g}\mathbf{d}\cdot\mathbf{E}_{\mathbf{\textbf{k}, \boldsymbol{\epsilon}}}^{(-)}(\mathbf{r}_{A})\ket{(N-1)_{\textbf{k}, \boldsymbol{\epsilon}}, 1_{\textbf{k}', \boldsymbol{\epsilon}'}; \;e, \;g} \bra{(N-1)_{\textbf{k}, \boldsymbol{\epsilon}}, 1_{\textbf{k}', \boldsymbol{\epsilon}'}; \;e, \;g}\nonumber\\
		&\times\mathbb{U}_{0}^{\dagger}(t-\tau) \; a_{\textbf{k}', \boldsymbol{\epsilon}'}^{\dagger}
		 a_{\textbf{k}', \boldsymbol{\epsilon}'} \; \mathbb{U}_{0}(t-\tilde{\tau})\ket{(N-1)_{\textbf{k}, \boldsymbol{\epsilon}}, 1_{\textbf{k}', \boldsymbol{\epsilon}'}; \;\tilde{e}, \;g}\bra{(N-1)_{\textbf{k}, \boldsymbol{\epsilon}}, 1_{\textbf{k}', \boldsymbol{\epsilon}'}; \;\tilde{e}, \;g} \mathbf{d}\cdot\mathbf{E}_{\mathbf{\textbf{k}', \boldsymbol{\epsilon}'}}^{(-)}(\mathbf{r}_{B}) \ket{(N-1)_{\textbf{k}, \boldsymbol{\epsilon}}; \;\tilde{e}, \;e}\\
		& \times \; \bra{(N-1)_{\textbf{k}, \boldsymbol{\epsilon}}; \;\tilde{e}, \;e} \mathbb{U}_{0}(\tilde{\tau}-\tilde{\tau}') e^{-\frac{\gamma}{2}(\tilde{\tau}-\tilde{\tau}')} \ket{(N-1)_{\textbf{k}, \boldsymbol{\epsilon}}; \;\tilde{e}, \;e}\bra{(N-1)_{\textbf{k}, \boldsymbol{\epsilon}}; \;\tilde{e}, \;e} \mathbf{d}\cdot\mathbf{E}_{\mathbf{\textbf{k}, \boldsymbol{\epsilon}}}^{(+)}(\mathbf{r}_{B})\ket{N_{\textbf{k}, \boldsymbol{\epsilon}}; \;\tilde{e}, \;g} \\
		& \times \; 
		\bra{N_{\textbf{k}, \boldsymbol{\epsilon}};\tilde{e}, g} \mathbb{U}_{0}(\tilde{\tau}')\ket{N_{\textbf{k}, \boldsymbol{\epsilon}}; \tilde{e}, g} \}
		= \frac{\mathcal{P}}{\Gamma}\;\frac{-\Omega_{0}^{2}}{4 \epsilon_{0} c^{2}}2\text{Re} \frac{d}{dt}  \int_{0}^{\infty} \frac{d\omega'}{\pi} \boldsymbol{\mu}\cdot \text{Im}\mathbb{G}(k' ;\textbf{R}) \cdot \boldsymbol{\mu}
		\frac{\omega'^{3}(e^{i\omega t} e^{-i\omega' t} + e^{-i\omega t} e^{i\omega' t})}{(\omega'-\omega_{0}+i\frac{\Gamma}{2})(\omega-\omega_{0}+i\frac{\gamma}{2})(\omega'-\omega)^{2}}\\
		 & = \frac{\Omega_{0}^{2} \omega}{\epsilon_{0} c^{2}} \frac{\mathcal{P}}{\Gamma}\; \frac{\omega^{2}\left[(\omega-\omega_{0})^{2}-\frac{\gamma \Gamma}{4}\right]  \boldsymbol{\mu} \cdot \text{Im}\mathbb{G}(\omega ;\textbf{R})\cdot \boldsymbol{\mu}}{[(\omega-\omega_{0})^{2}+\frac{\Gamma^{2}}{4}][(\omega-\omega_{0})^{2}+\frac{\gamma^{2}}{4}]},\qquad\gamma t\gtrsim1.\label{scat2expr}
	\end{split}
\end{equation}

As for the scattered power associated to the process of diagram (4) in Fig.\ref{fig2}, in which the emitted photons are created and annihilated at atom $B$, reads

\begin{equation}
	\begin{split}
		\mathcal{W}_{sc}^{(4)}
		%& =\frac{\mathcal{P}}{\Gamma} \frac{d}{dt}\textrm{Re} \;
		& =\frac{d}{dt}\;
		\langle\Psi_{4}(t)|\Psi_{4}^{f}\rangle\bra{\Psi_{4}^{f}}H_{EM}\ket{\Psi_{4}^{f}} \langle\Psi_{4}^{f}|\Psi_{4}(t)\rangle\\
		& = \frac{\mathcal{P}}{\Gamma}\;\frac{d}{dt} 2\textrm{Re}\sum_{\textbf{k}'', \boldsymbol{\epsilon}''} \hbar \omega'' \{ \hbar ^{-4} \int_{0}^{t} d\tau \int_{0}^{\tau} d \tau'  \int_{0}^{t} d\tilde{\tau} \int_{0}^{\tilde{\tau}} d \tilde{\tau}' \bra{N_{\mathbf{\textbf{k}, \boldsymbol{\epsilon}}}; \;\tilde{e}, \;g} \mathbb{U}_{0}^{\dagger}(\tau'')\ket{N_{\textbf{k}, \boldsymbol{\epsilon}}; \;\tilde{e}, \;g} \bra{N_{\textbf{k}, \boldsymbol{\epsilon}}; \;\tilde{e}, \;g} \\
		& \times \;  \mathbf{d}\cdot\mathbf{E}_{\mathbf{\textbf{k}', \boldsymbol{\epsilon}'}}^{(+)}(\mathbf{r}_{A}) \ket{N_{\textbf{k}, \boldsymbol{\epsilon}},1_{\textbf{k}', \boldsymbol{\epsilon}'} ; \;g, \;g}
		\bra{N_{\textbf{k}, \boldsymbol{\epsilon}},1_{\textbf{k}', \boldsymbol{\epsilon}'} ; \;g, \;g} e^{-\frac{\Gamma}{2}(\tau-\tau')} \mathbb{U}_{0}^{\dagger}(\tau'-\tau'')\ket{N_{\textbf{k}, \boldsymbol{\epsilon}}, 1_{\textbf{k}', \boldsymbol{\epsilon}'} ; \;g, \;g} \\
		& \times \; \bra{N_{\textbf{k}, \boldsymbol{\epsilon}},  1_{\textbf{k}', \boldsymbol{\epsilon}'} ; \;g, \;g}  \mathbf{d}\cdot\mathbf{E}_{\mathbf{\textbf{k}', \boldsymbol{\epsilon}'}}^{(-)}(\mathbf{r}_{B})\ket{N_{\textbf{k}, \boldsymbol{\epsilon}}; \;g, \;e}
		\bra{N_{\textbf{k}, \boldsymbol{\epsilon}}; \;g, \;e} e^{-\frac{\Gamma}{2}(\tau-\tau')}e^{-\frac{\gamma}{2}(\tau-\tau')} \mathbb{U}_{0}^{\dagger}(\tau-\tau') \\
		& \times \; \ket{N_{\textbf{k}, \boldsymbol{\epsilon}}; \;g, \;e} \bra{N_{\textbf{k}, \boldsymbol{\epsilon}}; \;g, \;e}  \mathbf{d}\cdot\mathbf{E}_{\mathbf{\textbf{k}, \boldsymbol{\epsilon}}}^{(-)}(\mathbf{r}_{A})\ket{(N-1)_{\textbf{k}, \boldsymbol{\epsilon}}; \;e, \;e}
		\bra{(N-1)_{\textbf{k}, \boldsymbol{\epsilon}}; \;e, \;e} e^{-\frac{\gamma}{2}(\tau-\tau')} \mathbb{U}_{0}^{\dagger}(\tau-\tau')  \\
		& \times \; \ket{(N-1)_{\textbf{k}, \boldsymbol{\epsilon}}; \;e, \;e} \bra{(N-1)_{\textbf{k}, \boldsymbol{\epsilon}}; \; e, \;e} \mathbf{d}\cdot\mathbf{E}_{\mathbf{\textbf{k}'', \boldsymbol{\epsilon}''}}^{(+)}(\mathbf{r}_{B}) \ket{(N-1)_{\textbf{k}, \boldsymbol{\epsilon}}, 1_{\textbf{k}'', \boldsymbol{\epsilon}''}; \;e, \;g} \\
		& \times \bra{(N-1)_{\textbf{k}, \boldsymbol{\epsilon}}, 1_{\textbf{k}'', \boldsymbol{\epsilon}''}; \;e, \;g}\mathbb{U}_{0}^{\dagger}(t-\tau) \; a_{\textbf{k}'', \boldsymbol{\epsilon}''}^{\dagger} \; a_{\textbf{k}'', \boldsymbol{\epsilon}''} \; \mathbb{U}_{0}(t-\tilde{\tau})\ket{(N-1)_{\textbf{k}, \boldsymbol{\epsilon}}, 1_{\textbf{k}'', \boldsymbol{\epsilon}''};\;\tilde{e}, \;g} \\
		& \times \bra{(N-1)_{\textbf{k}, \boldsymbol{\epsilon}}, 1_{\textbf{k}'', \boldsymbol{\epsilon}''};\;\tilde{e}, \;g}  \mathbf{d}\cdot\mathbf{E}_{\mathbf{\textbf{k}'', \boldsymbol{\epsilon}''}}^{(-)}(\mathbf{r}_{B}) \ket{(N-1)_{\textbf{k}, \boldsymbol{\epsilon}}; \;\tilde{e}, \;e}
		\; \bra{(N-1)_{\textbf{k}, \boldsymbol{\epsilon}}; \;\tilde{e}, \;e} \mathbb{U}_{0}(\tilde{\tau}-\tilde{\tau}') e^{-\frac{\gamma}{2}(\tilde{\tau}-\tilde{\tau}')}  \\
		& \times \; \ket{(N-1)_{\textbf{k}, \boldsymbol{\epsilon}}; \;\tilde{e}, \;e} \bra{(N-1)_{\textbf{k}, \boldsymbol{\epsilon}}; \;\tilde{e}, \;e} \mathbf{d}\cdot\mathbf{E}_{\mathbf{\textbf{k}, \boldsymbol{\epsilon}}}^{(+)}(\mathbf{r}_{B})\ket{N_{\textbf{k}, \boldsymbol{\epsilon}}; \;\tilde{e}, \;g}
		\bra{N_{\textbf{k}, \boldsymbol{\epsilon}}; \;\tilde{e}, \;g} \mathbb{U}_{0}(\tilde{\tau}')\ket{N_{\textbf{k}, \boldsymbol{\epsilon}}; \;\tilde{e}, \;g} \}\\
		& = \frac{\mathcal{P}}{\Gamma}\;\frac{-\Omega_{0}^{2}}{ 4\hbar \epsilon_{0}^{2} c^{4}}2\text{Re} \frac{d}{dt}  \int_{0}^{\infty} \frac{d\omega'}{\pi} \omega'^{2} \boldsymbol{\mu}\cdot \; \text{Im}\mathbb{G}(\omega' ;\textbf{R}) \; \cdot\boldsymbol{\mu} \int_{0}^{\infty} \frac{d\omega''}{\pi} \omega^{''3} \boldsymbol{\mu}\cdot \; \text{Im}\mathbb{G}(\omega'' ;\textbf{r}) \; \cdot\boldsymbol{\mu} \\
		& \times \frac{e^{-i\omega t} e^{i\omega'' t} + e^{i\omega t} e^{-i\omega'' t}}{(\omega'-\omega_{0}+i\frac{\Gamma}{2})(i\frac{\Gamma}{2}+i\frac{\gamma}{2})(\omega-\omega_{0}-i\frac{\gamma}{2})(\omega-\omega_{0}+i\frac{\gamma}{2})(\omega''-\omega)^{2}}\\
		& = \frac{\mathcal{P}}{\Gamma}\;\frac{\Omega_{0}^{2} \omega}{ \hbar \epsilon_{0}^{2} c^{4}} \; \frac{ - [\omega^{2} \; \boldsymbol{\mu}\cdot \; \text{Im}\mathbb{G}(\omega ;\textbf{r}) \cdot\boldsymbol{\mu}] [\omega_{0}^{2} \; \boldsymbol{\mu}\cdot \; \text{Im}\mathbb{G}(\omega_{0} ;\textbf{R}) \; \cdot\boldsymbol{\mu}]}{[(\omega-\omega_{0})^{2}+\frac{\gamma^{2}}{4}] \left( \frac{\Gamma}{2} + \frac{\gamma}{2} \right)},\qquad r\rightarrow 0^{+},\quad\gamma t\gtrsim1.
	\end{split}
\end{equation}

The expression corresponding to the diagram (9) of Fig.\ref{fig3} for the power absorbed collectively by the system along the 'transient' excitation of the passive atom from  state $\tilde{g}$ to $e$ at rate $\gamma$, reads

\begin{equation}
	\begin{split}
		\mathcal{W}^{(9)}
		& =\gamma \left[\bra{\Psi_{9}(0)}H_{EM}\ket{\Psi_{9}(0)}-
		\bra{\Psi_{9}(t)}\ket{\Psi_{9}^{f}} \bra{\Psi_{9}^{f}}H_{EM}\ket{\Psi_{9}^{f}} \bra{\Psi_{9}^{f}}\ket{\Psi_{9}(t)} \right] \\
		& = \gamma \; \frac{\gamma}{\Gamma}\; \sum_{\textbf{k}^{'}, \boldsymbol{\epsilon}^{'}} \bra{N_{\mathbf{\textbf{k}, \boldsymbol{\epsilon}}}; \;\tilde{g}, \;g}\hbar \omega' a_{\mathbf{\textbf{k}^{'}, \boldsymbol{\epsilon}^{'}}}^{\dagger} a_{\mathbf{\textbf{k}^{'}, \boldsymbol{\epsilon}^{'}}} \ket{N_{\mathbf{\textbf{k}, \boldsymbol{\epsilon}}}; \;\tilde{g}, \;g} \\
		& - \gamma \; \frac{\gamma}{\Gamma}\; \sum_{\textbf{k}^{'}, \boldsymbol{\epsilon}^{'}} \hbar \omega' \{ \hbar ^{-4} \int_{0}^{t} d\tau \int_{0}^{\tau} d \tau'  \int_{0}^{\tau'} d \tau''\int_{0}^{t} d\tilde{\tau} \bra{N_{\mathbf{\textbf{k}, \boldsymbol{\epsilon}}}; \;\tilde{g}, \;g} \mathbb{U}_{0}^{\dagger}(\tau'')\ket{N_{\textbf{k}, \boldsymbol{\epsilon}}; \;\tilde{g}, \;g} \\
		& \times \; \bra{N_{\textbf{k}, \boldsymbol{\epsilon}}; \;\tilde{g}, \;g} \mathbf{d}\cdot\mathbf{E}_{\mathbf{\textbf{k}^{'}, \boldsymbol{\epsilon}^{'}}}^{(-)}(\mathbf{r}_{A})\ket{(N-1)_{\textbf{k}, \boldsymbol{\epsilon}}; \;e, \;g} \bra{(N-1)_{\textbf{k}, \boldsymbol{\epsilon}}; \;e, \;g}  e^{-\frac{\Gamma}{2}(\tau'-\tau'')} \mathbb{U}_{0}^{\dagger}(\tau'-\tau'') \\
		& \times \; \ket{(N-1)_{\textbf{k}, \boldsymbol{\epsilon}}; \;e, \;g} \bra{(N-1)_{\textbf{k}, \boldsymbol{\epsilon}}; \;e, \;g}\mathbf{d}\cdot \mathbf{E}_{\mathbf{\textbf{k}^{'}, \boldsymbol{\epsilon}^{'}}}^{(+)}(\mathbf{r}_{A}) \ket{(N-1)_{\textbf{k}, \boldsymbol{\epsilon}}, 1_{\textbf{k}^{'}, \boldsymbol{\epsilon}^{'}};\;g, \;g} \\
		& \times \; \bra{(N-1)_{\textbf{k}, \boldsymbol{\epsilon}}, 1_{\textbf{k}^{'}, \boldsymbol{\epsilon}^{'}};\;g, \;g}\mathbb{U}_{0}^{\dagger}(\tau-\tau') \ket{(N-1)_{\textbf{k}, \boldsymbol{\epsilon}}, 1_{\textbf{k}^{'}, \boldsymbol{\epsilon}^{'}};\;g, \;g}  \bra{(N-1)_{\textbf{k}, \boldsymbol{\epsilon}},1_{\textbf{k}^{'}, \boldsymbol{\epsilon}^{'}}; \;g, \;g} \\
		& \times \;  \mathbf{d}\cdot \mathbf{E}_{\mathbf{\textbf{k}^{'}, \boldsymbol{\epsilon}^{'}}}^{(-)}(\mathbf{r}_{B}) \ket{(N-1)_{\textbf{k}, \boldsymbol{\epsilon}}; \;g, \;e} \bra{(N-1)_{\textbf{k}, \boldsymbol{\epsilon}}; \;\tilde{g}, \;e} e^{-\frac{\gamma}{2}(t-\tau)} \mathbb{U}_{0}^{\dagger}(t-\tau) \; a_{\mathbf{\textbf{k}^{'}, \boldsymbol{\epsilon}^{'}}}^{\dagger} a_{\mathbf{\textbf{k}^{'}, \boldsymbol{\epsilon}^{'}}}\mathbb{U}_{0}(t-\tilde{\tau}) e^{-\frac{\gamma}{2}(t-\tilde{\tau})}\\
		& \times \; \ket{(N-1)_{\textbf{k}, \boldsymbol{\epsilon}}; \;\tilde{g}, \;e} \bra{(N-1)_{\textbf{k}, \boldsymbol{\epsilon}}; \;\tilde{g}, \;e} \mathbf{d}\cdot \mathbf{E}_{\mathbf{\textbf{k}, \boldsymbol{\epsilon}}}^{(+)}(\mathbf{r}_{B})\ket{N_{\textbf{k}, \boldsymbol{\epsilon}}; \;\tilde{g}, \;g} \bra{N_{\textbf{k}, \boldsymbol{\epsilon}}; \;\tilde{g}, \;g} \mathbb{U}_{0}(\tilde{\tau})\ket{N_{\textbf{k}, \boldsymbol{\epsilon}}; \;\tilde{g}, \;g} \}\\
			& = \gamma\frac{\gamma}{\Gamma}\frac{\Omega_{0}^{2}}{ 4 \epsilon_{0} c^{2}}2\text{Re}\int_{0}^{\infty} \frac{d\omega'}{\pi} \boldsymbol{\mu}\cdot\text{Im}\mathbb{G}(\omega' ;\textbf{R})\cdot\boldsymbol{\mu}\frac{ \omega'^{3}[(\omega'-\omega_{0}-i\frac{\gamma}{2}) - (\omega-\omega_{0}-i\frac{\gamma}{2}) e^{-i\omega t} e^{i\omega' t}]}{(\omega-\omega_{0}-i\frac{\Gamma}{2})(\omega-\omega_{0}+i\frac{\gamma}{2})(\omega-\omega_{0}-i\frac{\gamma}{2})(\omega'-\omega_{0}-i\frac{\gamma}{2})(\omega'-\omega)}\\
		& = \gamma \; \frac{\gamma}{\Gamma}\;\frac{\Omega_{0}^{2} \omega^{3}}{4 \epsilon_{0} c^{2}} \frac{2(\omega - \omega_{0})  \boldsymbol{\mu}\cdot \text{Re}\mathbb{G}(\omega ;\textbf{R})\cdot\boldsymbol{\mu}+ \Gamma \boldsymbol{\mu}\cdot\text{Im}\mathbb{G}(\omega ;\textbf{R})\cdot \boldsymbol{\mu}}{[(\omega-\omega_{0})^{2}+\frac{\Gamma^{2}}{4}][(\omega-\omega_{0})^{2}+\frac{\gamma^{2}}{4}]},\quad\gamma t\gtrsim1.
	\end{split}
\end{equation}

As for the power emitted associated to the process of diagram (12) in Fig.\ref{fig3}, which corresponds to collective stimulated emission, it reads

\begin{equation}
	\begin{split}
		\mathcal{W}^{(12)}
		& =\Gamma \left[\bra{\Psi_{12}(0)}H_{EM}\ket{\Psi_{12}(0)}-
		\langle\Psi_{12}(t)|\Psi_{12}^{f}\rangle \bra{\Psi_{12}^{f}}H_{EM}\ket{\Psi_{12}^{f}} \langle\Psi_{12}^{f}|\Psi_{12}(t)\rangle\right] \\
		& =\Gamma \; \frac{\mathcal{P}}{\Gamma}\; \sum_{\textbf{k}', \boldsymbol{\epsilon}'} \bra{N_{\mathbf{\textbf{k}, \boldsymbol{\epsilon}}}; \;\tilde{e}, \;g}\hbar \omega' a_{\mathbf{\textbf{k}', \boldsymbol{\epsilon}'}}^{\dagger} a_{\mathbf{\textbf{k}', \boldsymbol{\epsilon}'}} \ket{N_{\mathbf{\textbf{k}, \boldsymbol{\epsilon}}}; \;\tilde{e}, \;g} \\
		& - \Gamma \; \frac{\mathcal{P}}{\Gamma}\; \sum_{\textbf{k}', \boldsymbol{\epsilon}'} \hbar \omega' \{ \hbar ^{-4} \int_{0}^{t} d\tau \int_{0}^{\tau} d \tau'  \int_{0}^{\tau'} d \tau''\int_{0}^{t} d\tilde{\tau} \bra{N_{\mathbf{\textbf{k}, \boldsymbol{\epsilon}}}; \;\tilde{e}, \;g} \mathbb{U}_{0}^{\dagger}(\tau'')\ket{N_{\textbf{k}, \boldsymbol{\epsilon}}; \;\tilde{e}, \;g} \\
		& \times \; \bra{N_{\textbf{k}, \boldsymbol{\epsilon}}; \;\tilde{e}, \;g} \mathbf{d.}\mathbf{E}_{\mathbf{\textbf{k}', \boldsymbol{\epsilon}'}}^{(+)}(\mathbf{r}_{A})\ket{N_{\textbf{k}, \boldsymbol{\epsilon}}; 1_{\textbf{k}', \boldsymbol{\epsilon}'}; \;g, \;g} \bra{N_{\textbf{k}, \boldsymbol{\epsilon}}; 1_{\textbf{k}', \boldsymbol{\epsilon}'}; \;g, \;g} e^{-\frac{\Gamma}{2}(\tau'-\tau'')} \mathbb{U}_{0}^{\dagger}(\tau'-\tau'') \\
		& \times \; \ket{N_{\textbf{k}, \boldsymbol{\epsilon}}; 1_{\textbf{k}', \boldsymbol{\epsilon}'}; \;g, \;g} \bra{N_{\textbf{k}, \boldsymbol{\epsilon}}; 1_{\textbf{k}', \boldsymbol{\epsilon}'}; \;g, \;g}\mathbf{d.}\mathbf{E}_{\mathbf{\textbf{k}', \boldsymbol{\epsilon}'}}^{(-)}(\mathbf{r}_{B}) \ket{N_{\textbf{k}, \boldsymbol{\epsilon}}; \;g, \;e} \bra{N_{\textbf{k}, \boldsymbol{\epsilon}}; \;g, \;e}  e^{-\frac{\Gamma}{2}} \mathbb{U}_{0}^{\dagger}(\tau-\tau') \\
		& \times \; e^{-\frac{\gamma}{2}} \ket{N_{\textbf{k}, \boldsymbol{\epsilon}}; \;g, \;e} \bra{N_{\textbf{k}, \boldsymbol{\epsilon}}; \;g, \;e} \mathbf{d.}\mathbf{E}_{\mathbf{\textbf{k}, \boldsymbol{\epsilon}}}^{(-)}(\mathbf{r}_{B}) \ket{(N+1)_{\textbf{k}, \boldsymbol{\epsilon}}; \;g, \;g} \bra{(N+1)_{\textbf{k}, \boldsymbol{\epsilon}}; \;g, \;g} e^{-\frac{\Gamma}{2}(t-\tau)} \\
		& \times \; \mathbb{U}_{0}^{\dagger}(t-\tau) a_{\textbf{k}', \boldsymbol{\epsilon}'}^{\dagger}\; a_{\textbf{k}', \boldsymbol{\epsilon}'} \; 
		\mathbb{U}_{0}(t-\tilde{\tau}) e^{-\frac{\Gamma}{2}(t-\tilde{\tau})} 	\ket{(N+1)_{\textbf{k}, \boldsymbol{\epsilon}}; \;g, \;g}
\bra{(N+1)_{\textbf{k}, \boldsymbol{\epsilon}}; \;g, \;g} \mathbf{d}\cdot\mathbf{E}_{\mathbf{\textbf{k}, \boldsymbol{\epsilon}}}^{(+)}(\mathbf{r}_{A})\\
		& \times \; \ket{N_{\textbf{k}, \boldsymbol{\epsilon}}; \;\tilde{e}, \;g} \bra{N_{\textbf{k}, \boldsymbol{\epsilon}}; \;\tilde{e}, \;g} \mathbb{U}_{0}(\tilde{\tau})\ket{N_{\textbf{k}, \boldsymbol{\epsilon}}; \;\tilde{e}, \;g} \}= \Gamma \; \frac{\mathcal{P}}{\Gamma}\;\frac{-\Omega_{0}^{2}}{ 4 \epsilon_{0} c^{2}}2\textrm{Re} \frac{d}{dt}  \int_{0}^{\infty} \frac{d\omega'}{\pi} \omega'^{3} \boldsymbol{\mu} \cdot\text{Im}\mathbb{G}(\omega' ;\textbf{R}) \cdot\boldsymbol{\mu} \\
		&  \times \; \frac{1}{(\omega'-\omega_{0}+i\frac{\Gamma}{2})(\omega-\omega_{0}+i\frac{\gamma}{2})(i\frac{\Gamma}{2}+i\frac{\Gamma}{2})} = \mathcal{P}\;\frac{\Omega_{0}^{2}}{  \epsilon_{0} c^{2}} \frac{\omega\omega_{0}^{2} \boldsymbol{\mu} \cdot \text{Im}\mathbb{G}(\omega_{0};\textbf{R}) \cdot \boldsymbol{\mu}}{[(\omega-\omega_{0})^{2}+\frac{\Gamma^{2}}{4}](\Gamma+\gamma)},\quad\gamma t\gtrsim1.
	\end{split}
\end{equation}

\section{Semiclassical calculation of the scattering cross-section}\label{appn}

Let us consider both atoms in their ground state in the absence of pump. Following Ref.\cite{myPRA}, the single atom polarizabilities for two-level atoms can be derived applying standard time-dependent perturbation theory to the computation of the complex-valued averaged expectation value of the dipole moment $\langle\mathbf{d}(\omega)\rangle$ induced by an external and quasi-resonant field of frequency $\omega$,
\begin{equation}
	\alpha(\omega)=\frac{\boldsymbol{\mu}\boldsymbol{\mu}}{\hbar(\omega-\omega_{0}+i\gamma/2)}.
\end{equation}
Applying classical linear response theory, the power emitted by each atom 
 is defined as the time-average rate of the interaction of
its induced dipole with the  electric field emitted by itself and its atomic partner, $\mathbf{E}^{*}(\mathbf{r}_{j},\omega)$,
\begin{equation}
\mathcal{W}_{sc}=\frac{-\omega}{2}\sum_{j=A,B}\textrm{Im}\:\{\langle\mathbf{d}_{j}(\omega)\rangle\cdot\mathbf{E}^{*}(\mathbf{r}_{j},\omega)\}=
\frac{-\omega \mu^{2} E_{0}^{2}}{\hbar} \; \text{Im} \; \left\{\frac{1-\frac{\omega^{2}}{\epsilon_{0} c}\alpha^{*}(\omega)\mathbb{G}^{*}(\textbf{R};\omega)}{\omega-\omega_{0}+i\gamma/2}\right\}.
\end{equation}
Replacing the dipole and EM field susceptibilities with their corresponding formulas and normalizing by the intensity of the external probe field, we obtain for the corresponding scattering cross-section,
\begin{equation}\label{equclas}
	\begin{split}
		\frac{\sigma_{sc}}{\sigma_{0}}
		& \simeq \left[\frac{\gamma \gamma_{0}}{2[(\omega-\omega_{0})^{2}+\frac{\gamma^{2}}{4}]}  + \frac{\left[(\omega-\omega_{0})^{2}-\frac{\gamma^{2}}{4}\right] \tilde{\Gamma}(R)\gamma_{0}  +  \left(\omega-\omega_{0}\right) \tilde{\Omega}(R)\gamma\gamma_{0}}{[(\omega-\omega_{0})^{2}+\frac{\gamma^{2}}{4}]^{2}}\right].
	\end{split}
\end{equation}

\end{widetext}


\begin{thebibliography}{105}
\bibitem{foot1} The contributions of diagrams (5) and (10) of Figs.\ref{fig2} and \ref{fig3}, for scattering and absorption respectively, are of an order $\gamma/\omega$ less than the leading order ones, and hence, negligible.
\end{thebibliography}
\end{document}